\documentclass[iop,apj,12pt,numberedappendix]{emulateapj}
\usepackage{times}
\usepackage[normalem]{ulem}
\usepackage{graphicx}
\usepackage{natbib}
\usepackage{amsfonts}
\usepackage{amsmath}
\usepackage{multirow}
\usepackage{enumerate}
\usepackage{comment}
\usepackage{float}
\usepackage{verbatim}
\usepackage[usenames,dvipsnames]{xcolor}
\usepackage[plainpages=false, colorlinks=true, anchorcolor=blue, linkcolor=blue, citecolor=blue, bookmarks=false]{hyperref}
\newcommand{\be}{\begin{eqnarray}}
\newcommand{\ee}{\end{eqnarray}}
\newcommand{\lp}{\left(}
\newcommand{\rp}{\right)}

\newcommand{\shortauth}{Morozova et al.}
\newcommand{\slugcom}{Accepted to ApJ, 10/15/2015}
\slugcomment{\slugcom}

\lefthead{\sc \footnotesize \slugcom \hfill \shortauth}
\righthead{\sc \footnotesize \slugcom \hfill \shortauth}

\begin{document}

\title{Light Curves of Core-Collapse Supernovae with Substantial Mass Loss\\
  using the New Open-Source SuperNova Explosion Code (SNEC)}

\author{Viktoriya Morozova\altaffilmark{1}}
\author{Anthony L. Piro\altaffilmark{2}}
\author{Mathieu Renzo\altaffilmark{3,1}}
\author{Christian D. Ott\altaffilmark{1}}
\author{Drew Clausen\altaffilmark{1}}
\author{Sean M. Couch\altaffilmark{1}}
\author{Justin Ellis\altaffilmark{1}}
\author{Luke F. Roberts\altaffilmark{1}}
\altaffiltext{1}{TAPIR, Walter Burke Institute for Theoretical Physics, MC 350-17,
  California Institute of Technology, Pasadena, CA 91125, USA, 
  morozvs@tapir.caltech.edu}
\altaffiltext{2}{Carnegie Observatories, 813 Santa Barbara Street, Pasadena, CA 91101, USA}
\altaffiltext{3}{Dipartimento di Fisica `Enrico Fermi,' University of Pisa, I-56127 Pisa, Italy}

\begin{abstract}
We present the SuperNova Explosion Code (\texttt{SNEC}), an
open-source Lagrangian code for the hydrodynamics and
equilibrium-diffusion radiation transport in the expanding envelopes
of supernovae. Given a model of a progenitor star, an explosion
energy, and an amount and distribution of radioactive nickel,
\texttt{SNEC} generates the bolometric light curve, as well as the
light curves in different broad bands assuming black body
emission. As a first application of \texttt{SNEC}, we consider the
explosions of a grid of $15\,M_\odot$ (at zero-age main sequence)
stars whose hydrogen envelopes are stripped to different extents and
at different points in their evolution. The resulting light curves
exhibit plateaus with durations of $\sim$$20-100\,\mathrm{days}$ if
$\gtrsim 1.5-2\,M_\odot$ of hydrogen-rich material is left and no
plateau if less hydrogen-rich material is left. If these shorter plateau
lengths are not seen for Type IIP supernovae in nature, it
suggests that, at least for zero-age main
sequence masses $\lesssim20\,M_\odot$, hydrogen mass loss occurs as an
all or nothing process. This perhaps points to the important role binary
interactions play in generating the observed mass-stripped supernovae (i.e.,
Type Ib/c events). These light curves are also unlike what is
typically seen for Type IIL supernovae, arguing that simply varying
the amount of mass loss cannot explain these events. The most stripped
models begin to show double-peaked light curves similar to what is
often seen for Type IIb supernovae, confirming previous work that
these supernovae can come from progenitors that have a small amount of
hydrogen and a radius of $\sim500\,R_\odot$.
\end{abstract}

  
\section{Introduction}

During the last half century the number of observed supernovae (SNe)
has increased exponentially \citep{minkowski:64,cappellaro:14}. Much
of this progress has been fueled by recent surveys, such as the Lick
Observatory Supernova Search \citep[LOSS,][]{leaman:11,smith:11}, the
Sloan Digital Sky Survey \citep[SDSS,][]{frieman:08}, and the Palomar
Transient Factory \citep[PTF,][]{rau:09}. In addition to providing
more complete and detailed samples of well-known classes of SNe (Type
Ia, Ib/c, II), these surveys have found a wide range of previously
unknown explosive events, from superluminous SNe \citep{quimby:11} to
rapid SN-like transients
\citep{perets:10,kasliwal:10,kasliwal:12,foley:13,inserra:15}. This
has opened our eyes to the broader range of astrophysical explosions
that can exist in nature.

Progress in explosive, transient observations has been closely
followed by progress in analytic and numerical light curve
modeling. For example, for SNe IIP, this has ranged from analytic
scalings \citep{arnett:80,chugai:91,popov:93} to detailed numerical
works
\citep[e.g.,][]{litvinova:83,chieffi:03,young:04,kasen:09,bersten:11,dessart:13}. These
investigations focused on understanding the general imprint of
progenitor characteristics (mass, radius, abundance and mixing of
$^{56}\mathrm{Ni}$, etc.)\ on the shape and luminosity of SN light
curves. In other cases, detailed comparisons have been made between
specific SNe II and numerical models
(\citealp{arnett:88,shigeyama:88,woosley:88,utrobin:93} for SN 1987A,
\citealp{nomoto:93,bartunov:94,shigeyama:94,young:95,blinnikov:98} for
SN 1993J, \citealp{baklanov:05,utrobin:07} for SN 1999em).

The combination of growing samples of SNe and other previously unknown
transients has motivated us to develop a new code for numerical
studies of the light curves of SN and SN-like explosions. Called the
SuperNova Explosion Code (\texttt{SNEC}), this general purpose code
will allow the user to take a stellar model (or other ad-hoc density
profile with other thermodynamic and compositional information), input
energy to generate an explosion, follow the hydrodynamic response, and
produce light curves.  The current iteration of \texttt{SNEC} is
spherically-symmetric (1D), and uses Lagrangian hydrodynamics and
equilibrium-diffusion (one-temperature) radiation transport. It also
follows other basic physics needed for light curves such as ionization
and heating from $^{56}\mathrm{Ni}$.  

In terms of complexity and amount of physics
  included, SNEC is at a somewhat intermediate position compared with
  existing SN light curve codes. The current state-of-the-art are
  multi-group radiation-hydrodynamic codes \citep[as
    in][]{blinnikov:93,moriya:11} and line-transfer radiative transfer
  codes that assume homologous expansion and either make the local
  thermodynamic equlibrium approximation (LTE; e.g.,
  \citealt{kasen:09}) or are fully non-LTE (e.g.,
  \citealt{hillier:12}). SNEC bridges the gap between these codes and
  analytical investigations, e.g., those of \cite{arnett:80},
  \cite{chugai:91}, and \cite{popov:93}, and the more recent ones of
  \cite{nakar:10}, \cite{goldfriend:14}, and \cite{nakar:15}. Our work
  is very much in the same spirit as the works of \citet{bersten:11}
  and \citet{ergon:15}.  A crucial aspect of \texttt{SNEC} is that
unlike these other codes, it is open source and publicly
available\footnote{\url{http://stellarcollapse.org/SNEC}}.  This will
make light curve modeling accessible and reproducible for the broader
community. It can be used for a wide range of studies, from generating
typical SN light curves as an educational tool to making light curves
for novel explosion scenarios. Modeling explosions and light curves
involves a wide range of physics and necessary approximations. Hence,
making code available and results reproducible is crucial for the
advancement of the field.

An important strength of \texttt{SNEC} is the ability to
systematically and quickly explore changes in stellar properties to
learn how they impact the resulting light curves. This is especially
useful for investigating the underlying mechanisms behind the
photometric diversity of SN II light curves, such as the SNe IIP
(with nearly constant plateau luminosity for a period $\sim 100$ days
past maximum, and the most common type), SNe IIL (with linearly
declining magnitude), and SNe~IIb (which show signatures of hydrogen
present, but with a light curve generally more similar to SNe
Ib)\footnote{We ignore SNe IIn for our study because they show clear
  signs of interaction, which is beyond the scope of this work.}.  In
particular, the connection between SNe IIP and SNe IIL has long been a
point of contention in the SN community. Early on, it was suggested by
\citet{barbon:79} and corroborated by \cite{blinnikov:93} that the
morphological differences might be explained by altering the envelope
masses while keeping the explosion mechanism the same (however, see
\citealp{swartz:91} for an alternative picture). In this explanation,
SNe IIL would simply have less hydrogen-rich envelopes than SNe
IIP. Nevertheless, SNe IIL must still have appreciable hydrogen
present, otherwise they would become SNe Ib/c instead. This suggests
that there may exist a continuous range of hydrogen mass stripping and
thus a continuous range of events between canonical SNe IIP and
IIL. Furthermore, SNe~IIb have been inferred to have a small amount of
hydrogen present ($\sim0.01-0.1\,M_\odot$,
\citealp{woosley:94,bersten:12,nakar:14}), and thus in principle with
sufficient mass loss a transition should be seen all the way to SNe
IIb. The question is whether additional ingredients are needed beyond
just increased mass loss to reproduce these features.

Motivated by these questions, we investigate the mass-loss hypothesis
for the origin of these SN II classes as a first
application of \texttt{SNEC}. We use presupernova stellar models
generated with the \texttt{MESA} code
\citep{paxton:11,paxton:13}. Besides providing excellent control in
generating models and investigating mass stripping, \texttt{MESA} has
been used for other light curve
studies \citep[e.g.,][]{dessart:13}, which allows us to compare
directly as a further check of \texttt{SNEC}.  Although
we find that varying levels of hydrogen mass stripping shortens the
plateau of the light curves, we conclude that simply varying the
amount of mass loss alone cannot explain the full range of properties
of SNe IIL.  In the most mass stripped cases, we begin to see
double-peaked light curves reminiscent of some SNe IIb, suggesting
that this transition occurs more naturally. Further work will be
needed for a more complete investigation of SNe IIb properties.

In Section \ref{SNEC}, we describe \texttt{SNEC} in detail. We follow,
in Section \ref{Progenitors}, with our study of massive stars with
varying levels of stripping. In Section \ref{Light_curves}, we present
the resulting SN light curves.  We conclude in Section
\ref{Conclusions} with a summary of our findings and a discussion of
future work. In Appendices \ref{Ap1} and
  \ref{Ap2}, we compare \texttt{SNEC} with two other codes.

\section{SNEC}
\label{SNEC}

We describe \texttt{SNEC} with a focus on SN IIP light curve modeling. Although
we anticipate that \texttt{SNEC} will continue to evolve and improve as it is
utilized for new projects and more physics is added, the discussion
below will provide some necessary background and summarize \texttt{SNEC}'s
general features. A more detailed description that also includes the
finite-difference form of the equations and implementation details is
available on the \texttt{\texttt{SNEC}} webpage,
\url{http://stellarcollapse.org/SNEC}.

It is also important to compare \texttt{SNEC}
with other SN light curve codes. Although below we focus on
what is implemented in SNEC, in Appendices \ref{Ap1} and \ref{Ap2}
we consider the work of \citet{bersten:11}, whose code has
a similar level of complexity as \texttt{SNEC}, and \citet{dessart:13},
whose code performs a more detailed treatment of the radiative transfer.
We find that both comparisons give satisfactory results with the main
difference being the transition from the plateau to the
$^{56}\mathrm{Ni}$ tail found by \citet{dessart:13}.
This disagreement likely reflects an intrinsic 
difference between equilibrium-diffusion radiation-hydrodynamics codes, such
as \texttt{SNEC}, and line-transfer radiative-transfer codes such as
\texttt{CMFGEN} used in \citet{dessart:13}.

\subsection{1D Lagrangian LTE Radiation Hydrodynamics with Ionization}

\label{sec:code}

Lagrangian hydrodynamics in spherical symmetry, supplemented with a
radiation diffusion term, written to the first order in $v/c$
\citep[see, e.g.,][]{mihalas:84,mezzacappa:93,bersten:10}, results
in a mass conservation equation (continuity equation),
\begin{equation}
\frac{\partial r}{\partial m} = \frac{1}{4\pi r^2 \rho}\ , \label{lag1}  
\end{equation}
an energy equation,
\begin{equation}
\frac{\partial \epsilon}{\partial t} = 
\frac{P}{\rho} \frac{\partial \ln \rho}{\partial t} 
- 4\pi r^2 Q \frac{\partial v}{\partial m} 
- \frac{\partial L}{\partial m} 
+ \epsilon_{\rm Ni} \,\, ,  \label{lag2}
\end{equation}
and a momentum equation,
\begin{equation}
\frac{\partial v}{\partial t} = 
-\frac{G m}{r^2} 
- 4 \pi r^2 \frac{\partial P}{\partial m} 
- 4 \pi \frac{\partial(r^2 Q)}{\partial m}\,\,.  \label{lag3}
\end{equation}
Here $m = \int_0^r 4\pi r'^2 \rho(r') dr'$ is the mass coordinate, $r$
is the radius, $t$ is the time, $\rho$ is the density, $\epsilon$ is
the specific internal energy, $P$ is the pressure, $v=\partial
r/\partial t$ is the velocity of the matter, $Q$ is the artificial
viscosity, and $G$ is the gravitational constant.
In Equation (\ref{lag2}), $\epsilon_{\rm Ni}$ is
  the specific energy deposition rate due to the radioactive decay of
  $^{56}\mathrm{Ni}$, which is equal to the time-dependent rate of
  energy release per gram of radioactive nickel $\epsilon_{\rm rad}$,
  multiplied by the deposition function $d$, both defined below in
  Section \ref{sec:nickel} by Equations (\ref{erad}) and (\ref{d}),
  respectively.  The radiative luminosity $L$ is
\begin{equation}
\label{luminosity}
L = - (4 \pi r^2)^2 \frac{\lambda a c}{3 \kappa}
\frac{\partial T^4}{\partial m} \ ,
\end{equation}
where $a$ is the radiation constant, $c$ is the speed of light,
$\lambda$ is the flux-limiter and $\kappa$ is the Rosseland mean
opacity. For capturing shocks, we use a simple von Neumann-Richtmyer
artificial viscosity \citep{vonneumann:50}
\begin{equation}
Q \equiv 
\begin{cases}
 C\rho(\partial v/\partial l)^2 
& \text{if }\ \partial v/\partial l < 0\,\,,\\
0              
& \text{otherwise}\,\, ,
\end{cases}
\end{equation}
where $l$ is an integer grid coordinate, and $C=2$. 
Following \cite{bersten:11}, we use the flux limiter
of \cite{levermore:81},
\begin{equation}
  \lambda = \frac{6 + 3R}{6+3R+R^2} \,\, ,
\end{equation}
where
\begin{equation}
R = \frac{4\pi r^2}{\kappa T^4} 
\left| \frac{\partial T^4}{\partial m} \right| \ .
\end{equation}

We discretize in mass and time following the scheme of
\citet{mezzacappa:93}. The mass conservation and momentum conservation
equations are updated time-explicitly. For Equation (\ref{lag2}), we
use a semi-implicit scheme with an adjustable parameter $\theta$ in
the discretization of the derivative ${\partial L}/{\partial m}$ that
can vary from fully explicit ($\theta=0$, only the luminosity from the
previous time step is used in the scheme) to fully implicit
($\theta=1$, only the luminosity at the next time step is used in the
scheme). The derivative $\partial T^4 / \partial m$ is linearized in
$\delta T$. We use $\theta=1/2$ for all simulations presented in this
paper. Using an initial guess for the temperature at the next time
step, we iteratively solve for $\delta T$, inverting a tridiagonal
matrix each time, until the fractional change in temperature is less
than a set tolerance ($10^{-7}$ in the current version of the
\texttt{SNEC}). We do not take the dependence of the opacity $\kappa$
on temperature into account in the implicit update and rather use the
opacity from the previous time step when solving for the temperature
at the next step.

\texttt{SNEC} assumes local thermodynamical equilibrium (LTE),
imposing the same temperature for radiation and matter. This
assumption is not valid at shock breakout and during and after the
transition phase from optically thick to optically thin ejecta. In SNe
IIP, it is reliable only during the plateau phase of the light curve
(see the discussions in \citealp{blinnikov:93,bersten:11}). However,
the comparison performed by \cite{bersten:10} between her LTE code
and a multi-group code suggests satisfactory agreement along the
entire light curve.

As a boundary condition for Equation (\ref{lag3}), we adopt $P=0$ 
at the center of the boundary cell, so half a
grid cell outside the star. For Equation (\ref{lag2}), we assume
that the luminosity at the surface is equal to the luminosity at the
closest interior grid point, i.e., that the diffusive term, $\partial
L/\partial m$, at the outer boundary is equal to zero. At the inner
boundary, we take the velocity and the bolometric luminosity to be
zero.  In the modeling of core-collapse SN light curves, the inner
boundary is typically not at $m=0$ due to the presence of a neutron
star (or a black hole), which is excluded from the grid. Setting the
inner velocity to zero excludes any possibility for fallback of
material onto the remnant in our models.

To close the system of hydrodynamic equations, we employ the analytic
equation of state (EOS) given by \citet{paczynski:83}, hereafter the
Paczy\'{n}ski EOS. The Paczy\'{n}ski EOS contains contributions from
radiation, ions, and electrons, and takes into account electron
degeneracy approximately. We repeat some of our model calculations
with the Helmholtz EOS \citep{timmes:99,timmes:00}, which includes a
(tabulated) complete electron EOS, to test the approximations made in
the Paczy\'{n}ski EOS.  We find that differences between the Paczy\'{n}ski and
Helmholtz EOS have negligible influence on the resulting light curves.

In order to account for recombination, we supplement the Paczy\'{n}ski EOS
with a routine that solves the Saha equations in the non-degenerate
approximation as proposed in \citet{zaghloul:00}. The set of Saha
equations, together with the condition of charge neutrality and number
conservation of nuclei of a given chemical element (enumerated by
index $k$), may be combined into a single transcendental equation for
the average charge $\bar{\mathbb{Z}}_k$ of element $k$ with atomic
number $Z_k$ as
\begin{equation}
  \label{saha}
  1-\bar{\mathbb{Z}}_k \left(\sum_{i=1}^{Z_k}\frac{i\displaystyle\prod_{j=1}^{i}f_{k,j}}{(\bar{\mathbb{Z}}_k n_k)^i}\right)^{-1}
        \left[1+\sum_{i=1}^{Z_k}\frac{\displaystyle\prod_{j=1}^{i}f_{k,j}}{(\bar{\mathbb{Z}}_k n_k)^i}\right] = 0\ ,
\end{equation}
with
\begin{equation}
  \begin{aligned}
  f_{k,i+1} = 2\frac{g_{k,i+1}}{g_{k,i}}\left[\frac{2\pi m_e k_{\rm B} T}{h^2}\right]^{3/2} \exp\left(-\frac{I_{k,i}}{k_{\rm B} T}\right) \,\, ,    \\ 
  i = 0,1,...,(Z_k-1) \,\, ,
  \end{aligned}
\end{equation}
where $i$ is the number of the ionization state ($0$ corresponds to
the neutral atom), ${Z}_k$ is the atomic number of element $k$, $n_k$
is the number density of element $k$, $g_{k,i}$ is the statistical
weight of the $i$-th ionization state of element $k$, $I_{k,i}$ is the
(positive) ionization energy for the ionization process
$i\rightarrow(i+1)$, $m_e$ is the electron rest mass, $h$ is Planck's
constant and $k_{\rm B}$ is Boltzmann's constant. Equation
(\ref{saha}) is solved iteratively at each call to the EOS, after
which the ionization fractions $\alpha_{k,i}$ are found as
\begin{equation}
\begin{aligned}
	\alpha_{k,0} &= \bar{\mathbb{Z}}_k \left(\sum_{i=1}^{\mathbb{Z}_k}\frac{i\displaystyle\prod_{j=1}^{i}f_{k,j}}{(\bar{\mathbb{Z}}_k n_k)^i}\right)^{-1} \ ,   \\
	\alpha_{k,i+1}&=\frac{\alpha_{k,i}}{\bar{\mathbb{Z}}_k n_k}f_{k,i+1}\ .
\end{aligned}
\end{equation}
Although one can consider as many elements as necessary at the expense
of computational time, for the present work we focus on the ionization
of hydrogen and helium. The specific internal energy is calculated as
\begin{equation}
	\epsilon = \epsilon_{\rm ion} + \epsilon_{\rm el} + \epsilon_{\rm rad} + \Delta\epsilon_{\rm ion}\ ,
\end{equation}
where $\epsilon_{\rm ion}$, $\epsilon_{\rm el}$, and $\epsilon_{\rm
  rad}$ are the contributions from ions, electrons, and radiation,
respectively, and
\begin{equation}
  \Delta
\epsilon_{\rm ion} = \frac{1}{\rho} \sum_k \left[ n_k
  \sum_{i=1}^{\mathbb{Z}_k} \lp \alpha_{k,i}\sum_{j=0}^{i-1}I_{k,j}
  \rp \right]
\end{equation}
is the ionization energy whose zero point for each element $k$ is the
neutral atom.

\subsection{Opacities}
\label{sec:opac}

The Rosseland mean opacity $\kappa$ is an essential input to our light
curve models. In the high temperature regime ($10^{3.75}\,\mathrm{K}<
T< 10^{8.7}\,\mathrm{K}$), we use the OPAL Type II opacity tables
\citep{iglesias:96} for solar metallicity ($Z_\odot = 0.02$
here). These tables allow for an increase in the mass fractions of two
chosen metals (in our case, carbon and oxygen) by deducting an amount
of helium to keep the sum of the mass fractions equal to unity. At low
temperatures ($10^{2.7}\,\mathrm{K} < T < 10^{4.5}\,\mathrm{K}$), we
use the tables of \citet{ferguson:05}. These tables are available for
solar composition, but not for enhanced carbon and oxygen mass
fractions compatible with the OPAL tables. In the overlap region
between the OPAL and the Ferguson~\emph{et al.}\ tables
($10^{3.75}\,\mathrm{K} < T < 10^{4.5}\,\mathrm{K}$), we give
preference to the low-temperature opacity tables, because they take
into account the contribution from molecular lines (this contribution
is not included in the OPAL tables). For carbon and oxygen enhanced
compositions, there are regions of low temperature and density for
which opacity values are not available. In these regions, the opacity
is generally most sensitive to temperature, and thus we set the
opacity to the nearest value that is available at the same
temperature. Most of the opacities set this way are below the opacity
floor (see the discussion below), so that this deficiency in the tables has a small
impact on the light curve evolution. At worst, it may affect the
transition between the plateau and the $^{56}\mathrm{Ni}$ dominated
part of SN IIP light curves when the photosphere first moves into
carbon/oxygen-rich regions.

The Rosseland mean opacity that we obtain from the OPAL and
Ferguson~\emph{et al.}\ tables does not describe all possible sources
of opacity needed for simulating SN light curves. As has been argued
in previous works
\citep[see][]{karp:77,young:04,blinnikov:96,bersten:11}, the tabulated
Rosseland mean opacity calculated for a static medium may
underestimate the contribution of the line opacities in the rapidly
expanding matter of the exploding star, plus it does not contain
possible non-thermal ionization and excitation by gamma rays. Due to
these missing effects, it is common practice to use a so-called
\emph{opacity floor}, effectively imposing a minimum possible value
for the opacity.  Presently, there is no universally agreed-upon
prescription for how to choose this opacity floor for a given
composition and velocity. In previous work, different values of the
opacity floor were chosen based either on simplified physical
arguments (e.g., \citealp{shigeyama:90}) or based on comparisons with
results obtained with multi-group or line-transfer codes (e.g.,
\citealp{bersten:11}). For SNe IIP, the values of the opacity floor
for the hydrogen-rich envelope and the metal-rich core of the star, as
well as the location and shape of the transition between the
core/envelope opacities, can strongly influence the shape of the
resulting light curve. Qualitatively (as shown in \citealp{bersten:10}
and confirmed by our simulations), a lower value of the opacity floor
in the envelope of the star increases the plateau luminosity and
decreases the duration of the plateau, and vice versa.  The luminosity
of the plateau and its duration are important observed photometric
quantities that are used for statistical studies of SNe IIP (as, e.g.,
in \citealp{anderson:14}). Hence, it is important to keep the
uncertainties in the opacity and their propagation into variations of
the light curve in mind when comparing modeling results with
observations.

In the work of \citet{bersten:11} on SN IIP light curves, the opacity
floor was set to $0.01\,\mathrm{cm}^2\,\mathrm{g}^{-1}$ for the
``envelope'' and $0.24\,\mathrm{cm}^2\,\mathrm{g}^{-1}$ for the
``core.'' Since \cite{bersten:11} are not specific in defining what
constitutes the core and envelope, and because this prescription introduces
large opacity discontinuities, we take a different approach in
\texttt{SNEC}. We choose the opacity floor to be linearly proportional
to metallicity $Z$ at each grid point, and set it to
$0.01\,\mathrm{cm}^2\,\mathrm{g}^{-1}$ for solar composition
($Z=0.02$) and to $0.24\,\mathrm{cm}^2\,\mathrm{g}^{-1}$ for a pure
metal composition ($Z=1$)\footnote{For the SNe IIP we study
  here, we do not find pure-metal regions in our models due to mixing
  that we impose during the explosion as described in Section
  \ref{sec:explosion_setup}.}. Note that we do not include the opacity floor in
the calculation of the optical depth and position of the photosphere
\citep[as in][]{bersten:11}. This is justified by the fact that the
opacity floor is used to account for line effects, which have minor
influence on the shape of most of the continuum spectrum. However, we
note that in the blue part of the spectrum (e.g., in the $U$ and $B$
bands), the continuum may be affected by the numerous
lines of iron-group elements (see, e.g., Figure 8 of
\citealp{kasen:09}).

\subsection{Radioactive $^{56}\mathrm{Ni}$ and Bolometric Luminosity}
\label{sec:nickel}

Radioactive $^{56}\mathrm{Ni}$ in core-collapse SNe is synthesized by
explosive nuclear burning of intermediate-mass elements during the
first seconds of the SN explosion in the inner regions of the star. It
is mixed outward by hydrodynamic instabilities triggered by the
shock's propagation through the envelope (see, e.g.,
\citealp{kifonidis:03,kifonidis:06,wongwathanarat:15}). The gamma
rays, emitted in the
$^{56}\mathrm{Ni}\rightarrow\,^{56}\mathrm{Co}\rightarrow\,^{56}\mathrm{Fe}$
decay process, diffuse and thermalize, providing an additional source of
energy $\epsilon_{\rm Ni}$ in Equation (\ref{lag2}). 

The present version of \texttt{SNEC} does not include a nuclear reaction
network and the amount and distribution of synthesized
$^{56}\mathrm{Ni}$ is provided by the user. While this is a technical
limitation that will be removed in future versions of \texttt{SNEC}, one should
keep in mind that nucleosynthetic yields are sensitive to (1) how the
explosion is launched, (2) where (in mass \emph{and} spatial
coordinate) it is launched, (3) to uncertainties in the structure and
composition of the layers in which explosive burning occurs (e.g.,
\citealt{young:07}). Specifying the $^{56}\mathrm{Ni}$ yield by hand
removes these uncertainties from our models and has the added benefit
of allowing  the user complete control of radioactive heating,
which can be useful for exploring how it impacts light curves.

For the gamma rays released in $^{56}\mathrm{Ni}$ and
$^{56}\mathrm{Co}$ decay, we follow the gray transfer approximation of
\citet{swartz:95}, solving the transfer equation in the form \be
\frac{dI'}{d\tau} = I' - X_{\rm Ni}\ , \ee where $\tau$ is the optical
depth along a given ray, $X_{\rm Ni}$ is the mass fraction of
$^{56}\mathrm{Ni}$, $I' = (4\pi\kappa_{\gamma}/\epsilon_{\rm rad})I$
and $I$ is the energy-integrated intensity. The effective gamma-ray
opacity is assumed to be purely absorptive and independent of energy,
$\kappa_{\gamma} = 0.06\,Y_{e}\,\mathrm{cm^2\,g^{-1}}$, where $Y_{e}$
is the electron fraction.  The time-dependent rate of energy release
per gram of radioactive nickel, $\epsilon_{\rm rad}$, is equal to \be
\label{erad}
&& \epsilon_{\rm rad} = 3.9\times 10^{10} \exp(-t/\tau_{\rm Ni}) +
6.78 \times 10^9 [ \exp(-t/\tau_{\rm Co}) \nonumber \\ &&
  \qquad\qquad\qquad - \exp(-t/\tau_{\rm Ni}) ]
\,\,\mathrm{erg\,g^{-1}\,s^{-1}}\ , \ee where $\tau_{\rm Ni}$ and
$\tau_{\rm Co}$ are the mean lifetimes of $^{56}\mathrm{Ni}$ and
$^{56}\mathrm{Co}$, equal to $8.8$ and $113.6$ days, respectively. The
local heating rate in each grid point is equal to $\epsilon_{\rm Ni} =
\epsilon_{\rm rad} d$, where $d$ is the deposition function \be
\label{d} d =
\frac{1}{4\pi}\oint I' d\omega, \ee where $\omega$ is the solid angle.

We do not take into account the energy from positrons released in the
radioactive decay of $^{56}\mathrm{Co}$ (which occurs for 19\% of the
decays). The total kinetic energy of positrons per $^{56}\mathrm{Co}$
decay is $\sim 0.12\,\mathrm{MeV}$ versus $\sim 3.61\,\mathrm{MeV}$
emitted via gamma rays \citep[see][]{nadyozhin:94}. Therefore,
neglecting this contribution constitutes an error of order $3-4\%$ in
the overall energetics of the $^{56}\mathrm{Ni}$ decay
\citep{swartz:95}.

Finally, we calculate the observed bolometric luminosity as suggested
in \citet{young:04}. It consists of two parts, the luminosity at the
photosphere and the luminosity due to the absorption of gamma rays
from $^{56}\mathrm{Ni}/^{56}\mathrm{Co}$ decay above the photosphere
\be L_{\rm obs} = L_{\rm photo} + \int_{m_{\rm photo}}^{M}S_{\rm
  dep}(m)dm\ .  \ee Here $m_{\rm photo}$ is the mass coordinate of the
photosphere, $M$ is the total mass of the star, $S_{\rm dep}$ is the
energy per gram per second deposited by gamma rays. The location of
the photosphere is defined by the optical depth $\tau = 2/3$, and
$L_{\rm photo}$ is found from Equation (\ref{luminosity}) at the
photosphere location.

\section{Progenitor models with Varying
  Hydrogen-rich Envelope Masses}
\label{Progenitors}
\label{MESA_progenitors}

\subsection{Motivation and Overall Strategy}
\label{sec:whenstrip}

As a first application of \texttt{SNEC} we investigate the light
curves of SNe from massive stars that have lost varying
amounts of their hydrogen-rich envelope during their evolution. We use
\texttt{SNEC} to explode presupernova models that we generate with the
open-source stellar evolution code \texttt{MESA} (release version
6794; \citealt{paxton:11,paxton:13}). The employed \texttt{MESA} input
files and final presupernova profiles are available at
\url{http://stellarcollapse.org/SNEC}.

Massive stars may lose large fractions of their hydrogen-rich
envelopes via steady line-driven winds, via stable or unstable binary
mass transfer, or, possibly, through pulsational instabilities and
eruptions (see, e.g., \citealt{smith:14} for a recent review). The
demographics of SN types combined with initial-mass-function
considerations suggest that line-driven winds alone cannot account for
the fraction of observed stripped-envelope SNe \citep{smith:11}. One
of the other avenues of mass loss may be required to partially or
completely remove hydrogen-rich envelope to account for the fraction
of observed Type IIb and Ib/c SNe. Since virtually all massive stars
are born in binaries and up to 70\% of them will interact with their
companion \citep{sana:12}, binary interaction may be the top contender
for how massive stars shed their hydrogen-rich envelopes. It is
possible that binary interactions (and other massergon:15-loss mechanisms) can
result in various degrees of envelope stripping and that there is a
broad distribution of hydrogen-rich envelope masses at the
presupernova stage of stars of any given zero-age main sequence (ZAMS)
mass.

Our goal here is to study the effects of substantial mass loss during
massive star evolution on presupernova structure and the resulting SN
light curve. We assume that the mass is lost rapidly (e.g., in an
unstable mass transfer event or through some instability), but
sufficiently slowly that the star can re-adjust to a new
equilibrium after the mass loss event. Rather than attempting to
self-consistently simulate various highly uncertain mass loss
mechanisms, we instead conduct a controlled experiment in stellar
astrophysics by systematically stripping material from the envelope of
a fiducial $M_\mathrm{ZAMS} = 15\,M_\odot$ star at different points of
its evolution. We note that \cite{bayless:15} carried out a similar
study of the effects of mass stripping on Type II SN light
curves. However, they considered a $23$-$M_\odot$ progenitor model
and stripped it only at the presupernova stage.

Figure~\ref{fig:when_stripping} shows the evolutionary track on the
Hertzsprung-Russel diagram for the $15$-$M_\odot$ reference
model. Rapid mass loss is most likely to occur in the post-MS
evolution because the envelope expands and becomes only weakly bound
to the compact core. When precisely the mass loss event occurs and how
much mass is lost will depend on the process causing mass loss and possibly
widely varying parameters such as the details of binary configuration.
In order to account for our ignorance of the details of the mass loss
event, we consider three points (indicated by symbols in
Figure~\ref{fig:when_stripping}; see also
Table~\ref{tab:stripped_setup}) in the post-MS evolution of our
reference model that probe different envelope structures and span
envelope radii from $\sim$$80 R_\odot$ to $\sim$$640 R_\odot$:

\begin{figure}[!t]
\centering
\includegraphics[width=0.48\textwidth]{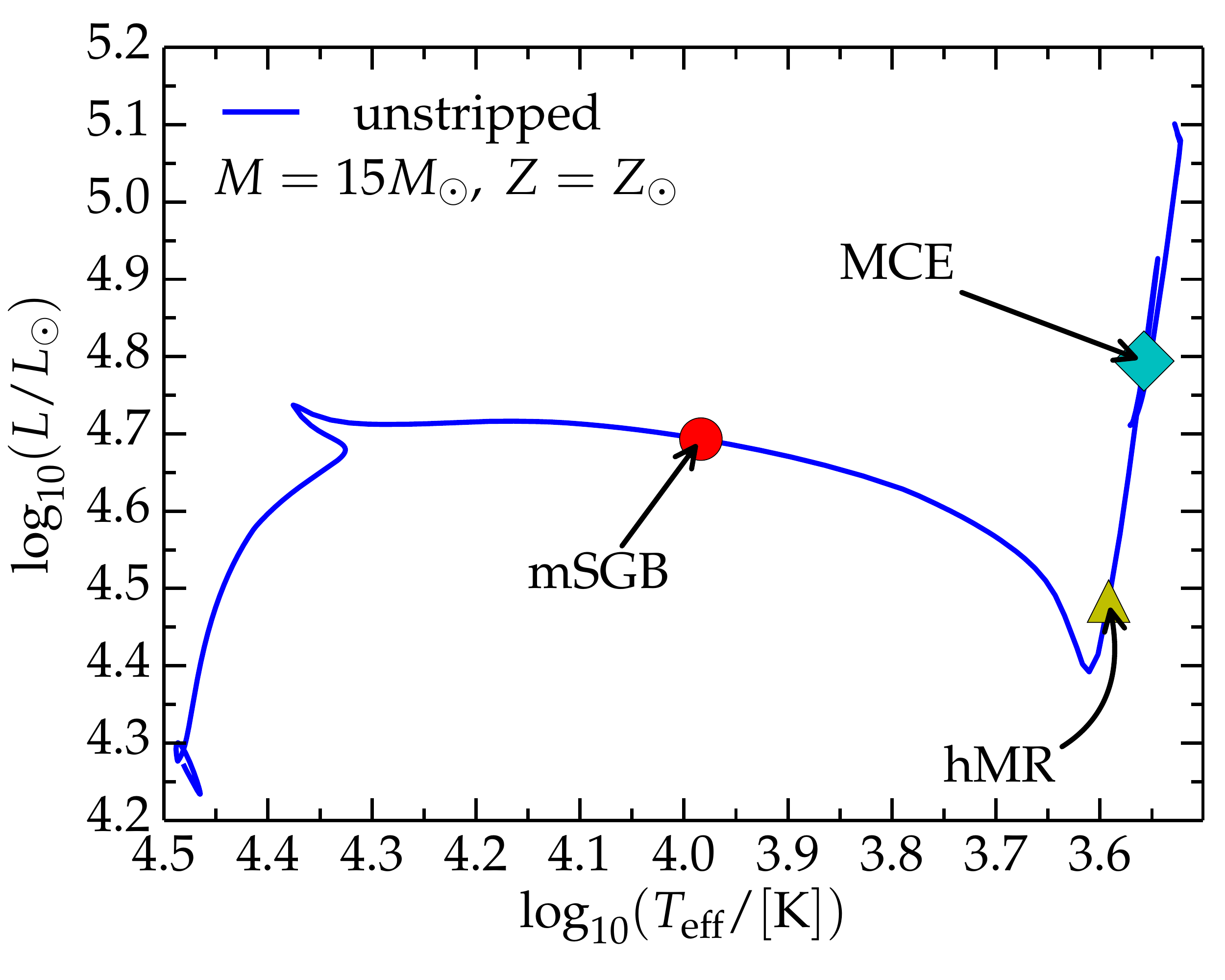}
\caption{Evolutionary track on the Hertzsprung-Russell diagram for the
  unstripped reference $15$-$M_\odot$ (at ZAMS) star computed with
  \texttt{MESA}. Each mark corresponds to a post-MS evolutionary stage
  at which we strip: mSGB (red dot) stands for ``middle of the
  subgiant branch'' defined via $T_\mathrm{eff} = 10^4\,\mathrm{K}$;
  hMR (yellow triangle) stands for ``half maximum radius,'' and MCE
  (cyan rhombus) stands for ``maximum extent of convective envelope.''
  While these three stripping points are separated only by
  $\sim$$10^4\,\mathrm{yrs}$ , they sample an interesting range of
  structure and an order of magnitude in envelope
  extent.  See Table~\ref{tab:stripped_setup} for more quantitative
  information on the stripping points.  Each stripped model
  corresponds to the reference model up to its stripping point and is
  stripped only once. \label{fig:when_stripping}}
\end{figure}

\begin{itemize}
\item \textbf{mSGB} series: These models are stripped at
  $T_\mathrm{eff} = 10^4\,\mathrm{K}$, which marks the middle of the
  subgiant branch (mSGB). At this point, the star's envelope has
  expanded to a radius of $79.8\,R_\odot$. Hydrogen is burning in
  a shell and the $\sim$$5\, M_\odot$ helium-rich core is inert. The
  envelope is still mostly radiative with a convective layer at mass
  coordinates $5.5 - 6.5\,M_\odot$.

\item \textbf{hMR} series: These models are stripped when the radius
  first surpasses the half-maximum radius (hMR) of the reference
  model, $R \sim 375\,R_\odot$. At this point, the envelope region
  outside a mass coordinate of $m \sim 9.5\,M_\odot$ is convective
  while deeper layers are radiative. There is 
  hydrogen shell burning and a small core helium burning region.
  
\item \textbf{MCE} series: These models are stripped when the maximum
  radial extent of the convective envelope is reached (MCE, at a
  stellar radius of $R\sim 638\,R_\odot$). The outer $\sim$$8 M_\odot$
  of the star are fully convective at this point.  There is hydrogen
  shell burning and a small core helium burning region.

\end{itemize}

\begin{table*}[!ht]
\renewcommand{\arraystretch}{1.3}
\caption{\raggedright Summary of stripping points.  We give the the
  criterion defining the stripping point, the unstripped reference
  model age, radius ($R$), hydrogen-rich envelope mass
  ($M_\mathrm{H}$), helium-core mass ($M_\mathrm{He}$), and the
  maximum mass stripped ($\mathrm{max}\{\Delta M\}$) for each
  stripping point. We generate stripped models at each stripping point
  at the timestep at which the reference model exactly meets or
  exceeds the stripping criterion. In the criterion for MCE, $X_c$ is
  the abundance of hydrogen in the central computational cell;
  $v_\mathrm{conv}^\mathrm{surf}$ is the unweighted average convective
  velocity in the outermost 150 computational cells,
  $v_\mathrm{conv}^\mathrm{env}$ is the unweighted average convective
  velocity in the 150 computational cells above the outermost lower
  boundary of a convective region. If these two differ by less than
  $10\%$, the envelope has roughly reached its maximum extent.}
    \label{tab:stripped_setup}
\centering
    \begin{tabular}{llccccc}\hline\hline
      Series Name & Stripping Criterion & Age [Myr] & $R \ [R_\odot]$ &
      $M_\mathrm{H} \ [M_\odot]$ & $M_\mathrm{He} \ [M_\odot]$ &
      $\max\{\Delta M\} $     \\\hline
      Middle SGB ({\bf mSGB}) & $T_\mathrm{eff}=10^4 \ \mathrm{K}$ &
      $13.0263$
      & \phantom{3}79.8 & 10.67 & 3.81 &7 \\
      Half Maximum Radius ({\bf hMR}) & $R\gtrsim375\ R_\odot$ &
      $13.0305$
      & 381.6 & 10.62 & 3.87 & 7 \\
      Maximum Extent  & $ X_c = 0$ and & $13.0310$
      & 638.1 & 10.61 & 3.88 & 7 \\[-0.2em] 
      of the Convective Envelope ({\bf MCE})& 
     $
     (v_\mathrm{conv}^\mathrm{surf} -
      v_\mathrm{conv}^\mathrm{env})/v_\mathrm{conv}^\mathrm{surf} \leq
      0.1$&&&&\\
\hline
    \end{tabular}  
\end{table*}

In each series of models, we strip in units of $1\,M_\odot$, but stop
before we reach the hydrogen shell burning region and keep at least $1
M_\odot$ of the radiative layer that surrounds it. Note that if we
considered Roche-lobe overflow in a binary system as the mechanism for
mass loss, then only the outer convective layers could be lost in an
unstable Roch-lobe overflow event, but stripping of radiative layers
would not occur dynamically (cf.~\citealt{hilditch:01}).

The time spanned by mSGB--hMR--MCE is only of order
$10^4\,\mathrm{yrs}$, which is small compared to the full lifetime of
the unstripped reference model ($\sim$$14.13 \times 10^6\,{\rm
  yrs}$). It is, however, very large compared to the dynamical time
and the thermal Kelvin-Helmholtz time of the star. The latter is
$t_\mathrm{KH} \sim 3/4\, G M^2 / (R L)$ \citep{kippenhahn:13}, which
is $\sim$$1250\,\mathrm{yrs}$, $\sim$$425\,\mathrm{yrs}$, and
$\sim$$125\,\mathrm{yrs}$ at mSGB, hMR, and MCE, respectively. After
stripping, about $\sim$$10^6\,\mathrm{yrs}$ of evolution are left
until core collapse.

\subsection{\texttt{MESA} Simulations}

We employ \texttt{MESA} release version 6794
\citep{paxton:11,paxton:13} and assume solar metallicity $Z_\odot =
0.019$. We use the Ledoux criterion for convection and follow
\cite{sukhbold:14}, who set the mixing
length parameter $\alpha_\mathrm{mlt}=2.0$, the overshooting parameter
$f_\mathrm{ov}=0.025$, and semiconvection efficiency
$\alpha_\mathrm{sc} = 0.01$, and do not consider thermohaline mixing.
We use the wind mass loss prescription of \cite{vink:00,vink:01} for
the hot MS phase and that of \cite{dejager:88} for the cool giant
phase, both with $\eta=1.0$. We limit \texttt{MESA}'s timestep by
enforcing fractional changes in structure and thermodynamics variables
of less than $10^{-3}$ per timestep
(\texttt{varcontrol\_target}$=10^{-3}$) and also use a
customized timestep control that enforces a timestep that is always
smaller than the model's Kelvin-Helmholtz time. We use \texttt{MESA}'s
standard setting for rezoning, \texttt{mesh\_delta\_coeff} $= 1.0$,
and \texttt{mesh\_delta\_coeff\_for\_highT} $=1.5$, which coarsens
the resolution at $T\gtrsim 10^9\,\mathrm{K}$ and, thus, in the core
region, where we do not currently trust \texttt{MESA} results (see below).
These standard resolution setting provide a sufficiently resolved
envelope for our \texttt{SNEC} explosion and light curve simulations.

For simplicity and speed of execution, we
simulate all models with \texttt{MESA}'s  default 21-isotope
nuclear reaction network \texttt{approx21} until the onset of core
collapse, which is commonly defined as the point when the infall
velocity reaches $1000\,\mathrm{km}\,\mathrm{s}^{-1}$. We note that
much larger ($100-1000$ isotope) networks are needed for an accurate
treatment of late oxygen burning and silicon burning and of the
pre-collapse neutronization in the degenerate core (e.g.,
\citealt{sukhbold:14}; Arnett, \emph{private communication}).  Since
the treatment of these late burning stages has a large effect on the
core region out to the carbon-oxygen -- helium interface
\citep{sukhbold:14}, the core structure of our \texttt{MESA} models is
not reliable.  Core collapse and postbounce supernova simulations that
focus on the explosion mechanism suggest that the structure in the
inner $1.4-2.5\,M_\odot$ may determine if neutrino-driven explosions
fail or succeed \citep{oconnor:11,ugliano:12,ertl:15}. However, for
our explosion study with \texttt{SNEC}, details of the core structure
are not essential, since the outer regions and the hydrogen-rich
envelope determine the light curve. We also artificially launch
explosions and introduce $^{56}$Ni by hand (see Section \ref{SNEC}).

\vspace*{0.5cm}
\subsection{Stripping Procedure}
\label{stripping_process}

We first evolve copies of the unstripped reference model to the onset
of core collapse, and to the three stripping points: mSGB, hMR, and
MCE (see Table~\ref{tab:stripped_setup} and
Fig.~\ref{fig:when_stripping}).  At these stripping points, we then
restart and use \texttt{MESA}'s module \texttt{adjust\_mass} to
``instantaneously'' remove the specified amount of mass. The new
smaller value of the total mass is reached using 80 \texttt{MESA}
``pseudo-evolution'' steps (i.e., the structure is evolved using
timesteps, but the time coordinate is held constant). During each
step, \texttt{MESA} removes the largest amount of mass from the
envelope that it can while still finding a hydrostatic solution to the
structure equations. In most cases, $\sim$$75$ are sufficient to reach
the desired mass and the last $\sim$$5$ have mass loss set to
zero. Both during and at the end of the pseudo-evolution, the
structure is always in global hydrostatic equilibrium, therefore, when
the regular evolution resumes, no readjustment occurs.

We strip mass in $1\,M_\odot$ steps and continue the
evolution of each model to the onset of core collapse. We refer to the
unstripped reference model simply as ``unstripped,'' and name the
stripped models according to [stripping point]\_[number of $M_\odot$
  stripped]M. For example, ``hMR\_5M'' stands for a model that had
$5\,M_\odot$ stripped at hMR.

\begin{figure}[t]
\centering
\includegraphics[width=.47\textwidth]{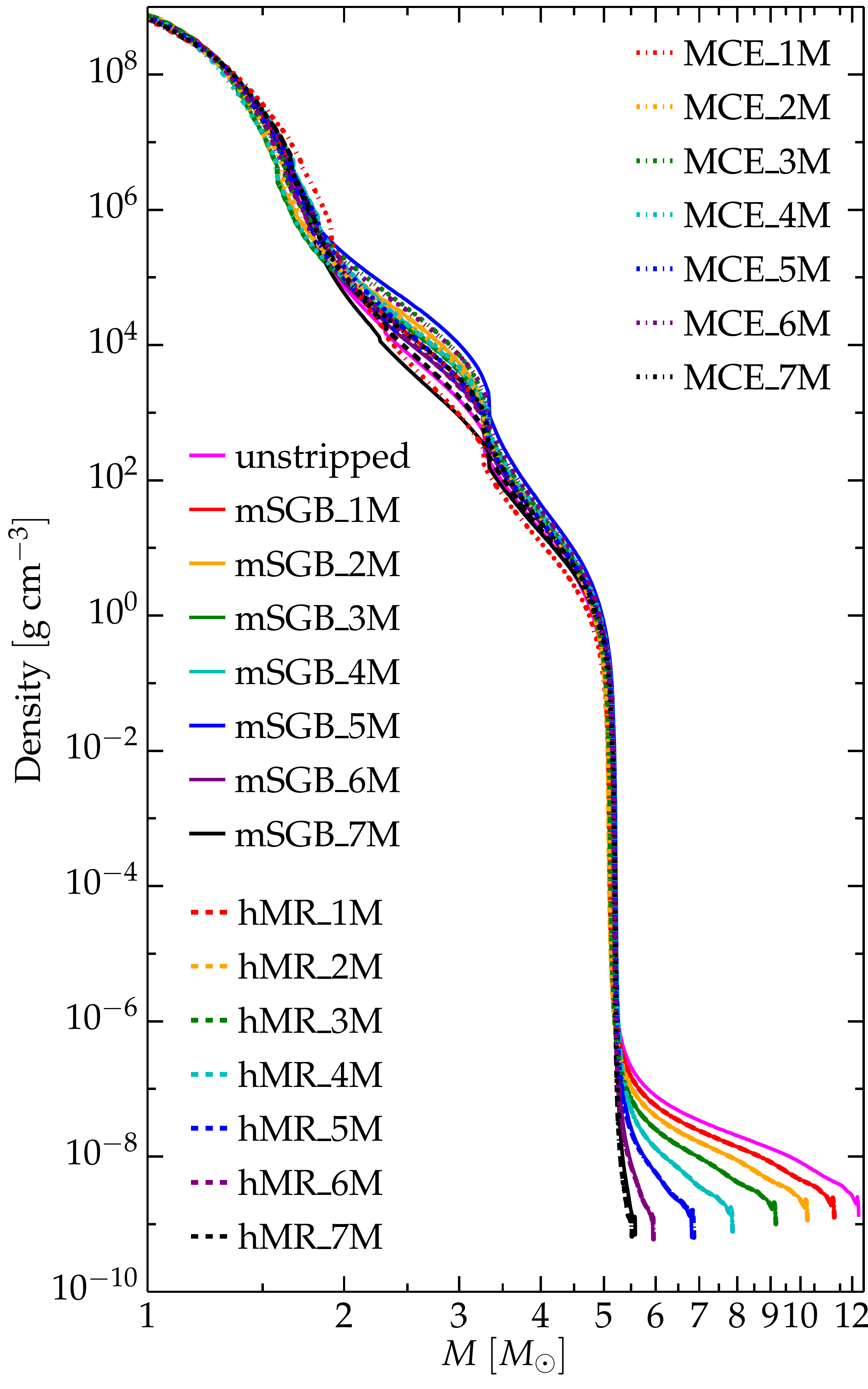}
\caption{Density profiles as a function of enclosed mass at the onset
  of core collapse of the entire set of \texttt{MESA} presupernova
  models computed in this study. The solid curves include the
  unstripped reference model and the models stripped at mSGB, dashed
  curves show models stripped at hMR, and dot-dashed curves represent
  models stripped at MCE (cf.~Figure~\ref{fig:when_stripping} and the
  text for a discussion of these stripping points). More stripped
  models have more tenuous envelopes, but the time of stripping has
  negligible influence on the envelope structure. However, both time
  of stripping and the amount of mass stripped influence the structure
  of the silicon and carbon/oxygen layers around $1.5 - 3.3\,M_\odot$
  in a complex and not obviously systematic way. The structure in this
  region may determine the outcome of core collapse
  (cf.~\citealt{oconnor:11,ugliano:12,ertl:15}).
\label{fig:rho_M}}
\end{figure}

\begin{table*}
\renewcommand{\arraystretch}{1.3}
\centering
\caption{\raggedright Summary of the presupernova structure of the
  \texttt{MESA} models.  $M_\mathrm{pre-SN}$ is the total presupernova
  mass, $M_\mathrm{H}$ is the mass of the hydrogen-rich envelope, and
  $M_\mathrm{He}$, $M_\mathrm{CO}$, and $M_\mathrm{Fe}$ are the
  helium, carbon/oxygen, and iron core masses, respectively.  We place
  the iron core boundary at the first location going inward where the
  electron fraction $Y_e < 0.49$, following the definition of
  \cite{dessart:13}.  $E_\mathrm{b}$ is the binding energy of the
  material outside $1.4\,M_\odot$ (the mass coordinate of the inner
  boundary in our \texttt{SNEC} explosion models) given in units of
  $\mathrm{B}$\emph{ethe}, $1\,\mathrm{B} = 10^{51}\,\mathrm{erg}$.
  $\xi_{2.5}^\mathrm{pre-SN}$ is the compactness parameter of
  \cite{oconnor:11} and $R$, $L$, and $T_\mathrm{eff}$ are the
  presupernova stellar radius, luminosity, and effective
  temperature. $t_\mathrm{SB}$ is the time of shock breakout in
  hours after the onset of the thermal bomb. \label{tab:stripped_res}}
\begin{tabular}{lccccccccccc}\hline \hline
Model & $M_\mathrm{pre-SN} \ [M_\odot]$ & $M_\mathrm{H} \ [M_\odot]$ & $M_\mathrm{He} \ [M_\odot]$
& $M_\mathrm{CO} \ [M_\odot]$ & $M_\mathrm{Fe} \ [M_\odot]$& $|E_{\mathrm{b}}|$ [B]&
$\xi_{2.5}^\mathrm{pre-SN}$ & $R \ [R_\odot]$ & $L \ [L_\odot]$ &
$T_\mathrm{eff} \ \mathrm{[10^3\ K]}$&$t_\mathrm{SB}\,[\mathrm{h}]$\\
\hline
unstripped & 12.28 & \phantom{$\lesssim$}7.18 & 5.10 & 3.27 & 1.51 & 0.641 & 0.103 & 1039 & 120309 & 3.337& 48.52\\ 
\hline
mSGB\_1M & 11.27 &\phantom{$\lesssim$}6.18 & 5.09 & 3.28 & 1.49 & 0.697 &
0.125 & 1031 & 121084 &3.355&45.71\\ 

mSGB\_2M & 10.25 &\phantom{$\lesssim$}5.16 & 5.09 & 3.26 & 1.49 & 0.617 &
0.142 & 1013 & 119370 & 3.373&42.79\\ 

mSGB\_3M & \phantom{0}9.17 &\phantom{$\lesssim$}4.06 & 5.11 & 3.27 &
1.49 & 0.586 & 0.127 & \phantom{0}991 & 121536 & 3.425&38.74\\ 

mSGB\_4M & \phantom{0}7.87 & \phantom{$\lesssim$}2.67 &5.20 & 3.32 &
1.58 & 0.749 & 0.138 & \phantom{0}932 & 122270 & 3.537&31.99\\ 

mSGB\_5M & \phantom{0}6.82 &\phantom{$\lesssim$}1.61 & 5.21 & 3.33 &
1.56 & 0.734 & 0.171 & \phantom{0}828 & 122984 & 3.759&25.11\\ 

mSGB\_6M & \phantom{0}5.94 & \phantom{$\lesssim$}0.74 &5.20 & 3.32 &
1.54 & 0.650 & 0.114 & \phantom{0}663 & 123258 & 4.204&16.75 \\ 

mSGB\_7M & \phantom{0}5.59 &\phantom{$\lesssim$}0.38 & 5.21 & 3.33 &
1.50 & 0.625 & 0.089 & \phantom{0}555 & 118763 & 4.553&12.25\\ 
\hline
hMR\_1M & 11.27 & \phantom{$\lesssim$}6.18 &5.09 & 3.28 & 1.49&0.697 & 0.125
& 1031 & 121084 & 3.355&45.71 \\ 

hMR\_2M & 10.25 & \phantom{$\lesssim$}5.16 &5.09 & 3.26 & 1.49&0.617 & 0.142
& 1013 & 119370 &3.373&42.79 \\ 

hMR\_3M & \phantom{0}9.17 & \phantom{$\lesssim$}4.06 & 5.11 & 3.27 & 1.49&0.586& 0.127 & \phantom{0}991 & 121536&3.425&38.74\\ 

hMR\_4M & \phantom{0}7.87 &  \phantom{$\lesssim$}2.67 & 5.20 & 3.32 & 1.58&0.749 & 0.138 & \phantom{0}932 & 122270&3.537&31.99\\ 

hMR\_5M & \phantom{0}6.87 &  \phantom{$\lesssim$}1.68 & 5.19 & 3.32 & 1.53&0.658 & 0.118 & \phantom{0}843 & 122179&3.719&25.58\\ 

hMR\_6M & \phantom{0}5.96 & \phantom{$\lesssim$}0.77 & 5.18 & 3.31 & 1.58 & 0.709 & 0.122 & \phantom{0}676 & 122065&4.153&17.19\\

hMR\_7M & \phantom{0}5.52 & \phantom{$\lesssim$}0.32 & 5.21 & 3.30 & 1.60 & 0.706 & 0.110 & \phantom{0}551 & 122645&4.604&11.83\\
\hline 

MCE\_1M & 11.27 &\phantom{$\lesssim$}6.17 & 5.10 & 3.27 & 1.58 & 0.765 & 0.102 & 1032 & 118857&3.339&45.62\\ 

MCE\_2M & 10.25 &\phantom{$\lesssim$}5.16 & 5.09 & 3.27 & 1.54& 0.644 & 0.134 & 1016 & 120982&3.379&42.79\\ 

MCE\_3M & \phantom{0}9.17 &\phantom{$\lesssim$}4.06 & 5.11 & 3.27 & 1.53 & 0.696 & 0.159 & \phantom{0}989 & 119197&3.413&38.65\\ 

MCE\_4M & \phantom{0}7.88 &\phantom{$\lesssim$}2.69 & 5.19 & 3.32 & 1.58 & 0.592 & 0.130 & \phantom{0}932 & 122808&3.541&32.36\\ 

MCE\_5M & \phantom{0}6.87 &\phantom{$\lesssim$}1.68 &  5.19 & 3.32 & 1.52 & 0.708 & 0.131 & \phantom{0}843 & 121709&3.715&25.55\\ 

MCE\_6M & \phantom{0}5.96 & \phantom{$\lesssim$}0.78 &5.18 & 3.31 & 1.55 &0.694 & 0.153 & \phantom{0}675 & 122791&4.162&17.20\\ 

MCE\_7M & \phantom{0}5.52 & \phantom{$\lesssim$}0.31 &5.21 & 3.31 & 1.56 & 0.718 & 0.123 & \phantom{0}552 & 122631&4.602&11.80\\

\hline
\end{tabular}
\end{table*}

\subsection{Resulting Presupernova Structure}
\label{sec:presnstructure}

We summarize the presupernova structure of our model set in
Table~\ref{tab:stripped_res} and Figure~\ref{fig:rho_M}. The
unstripped reference model reaches the presupernova stage as a
$12.28\,M_\odot$ RSG with a hydrogen-rich envelope
mass of $M_\mathrm{H} \sim 7.18\,M_\odot$. The stripped models have
envelopes of systematically lower mass, approximately proportional to
the amount of mass removed. While $M_\mathrm{H}$ varies by more than a
factor of seven ($M_\mathrm{H} \sim 0.31-0.38$ for the most stripped
models) within a model series, the final stellar radius varies only by
a factor of $\sim$$2$. Hence, the envelopes become increasingly dilute
(lower-density) with increasing stripping. Following the
$T_\mathrm{eff}$ criterion of \cite{georgy:12}, our models with less
than $6\,M_\odot$ stripped die as RSGs while models from which we
strip $6\,M_\odot$ or $7\, M_\odot$ die as yellow supergiants
(YSGs). It is apparent from Figure~\ref{fig:rho_M} that the choice of
stripping point has almost negligible influence on the final mass and
structure of the remaining hydrogen-rich envelope.  Differences in
$M_\mathrm{H}$ and radius $R$ are generally $\lesssim 5\%$ between
models from different series that had the same amount of mass
removed. An exception are the most extreme \_M7 cases ($7 M_\odot$ of
hydrogen-rich material stripped) that show a $\sim20\%$ variation in
their final $M_\mathrm{H}$ from $0.31 M_\odot$ in model MCE\_7M, $0.32
M_\odot$ in model hMR\_7M, to $0.38\,M_\odot$ in model mSGB\_7M. The
envelope in the latter model temporarily becomes compact when helium
ignites in the core, which leads to less wind mass loss after
stripping. The radii of all \_7M models at core collapse are nearly
identical ($\sim$$550\,R_\odot$).

The iron core mass ($\sim$$1.5-1.6\,M_\odot$) and density profile is
nearly identical in all models. They reach core collapse at central
densities in the range $0.93 - 1.48\times
10^{10}\,\mathrm{g\,cm}^{-3}$.  Since the electrons in the iron core
are relativistically degenerate and the core specific entropy and
electron fraction are roughly the same in all models, the iron core
structure is very similar throughout the model series. More
interesting are the large variations in the density profiles in the
silicon and oxygen/carbon layers above the iron core, between
$\sim$$1.5\,M_\odot$ and $\sim$$3.3\,M_\odot$, as shown in
Figure~\ref{fig:rho_M}. The presupernova structure in these layers
appears to be very sensitive to both the amount of envelope mass
stripped and the stripping point in the evolution. However, there are
no identifiable trends that could be linked to amount of mass
stripped and stripping point. \cite{sukhbold:14} pointed out that the
structure in these layers is sensitive to the treatment of (1)
nuclear reactions and weak interactions (neutrino cooling,
neutronization) and (2) mixing and overshooting. Both
(1) and (2), in turn, influence the number and extent of convective
shell burning episodes/regions in the silicon and carbon/oxygen
layers. Our results indicate that variations in the time and amount of
mass loss can also influence this part of presupernova stellar
structure. The density distribution in the affected regions determines
the compactness parameter of \cite{oconnor:11},
\begin{equation}
\xi_{M}
\stackrel{\mathrm{def}}{=} \frac{M/M_\odot}{R(M)/1000 \ \mathrm{km}}
\ \ ,
\end{equation}
with the commonly adopted reference value $M=2.5M_\odot$. Multiple
studies (e.g., \citealt{ertl:15,ugliano:12,oconnor:11}) have demonstrated
that the compactness parameter $\xi_{2.5}$ is a useful quantity to (at a roughly
``first order'' level) judge whether a given star is more likely to explode
in a supernova or collapse to a black hole without explosion.
Hence, the dependence of $\xi_{2.5}$ on mass loss (both rapid and due to winds)
 deserves further investigation in future work with a version of
\texttt{MESA} with a much larger nuclear reaction network and a more
reliable treatment of the final stages of stellar evolution.

\begin{figure}
\centering
\includegraphics[width=0.475\textwidth]{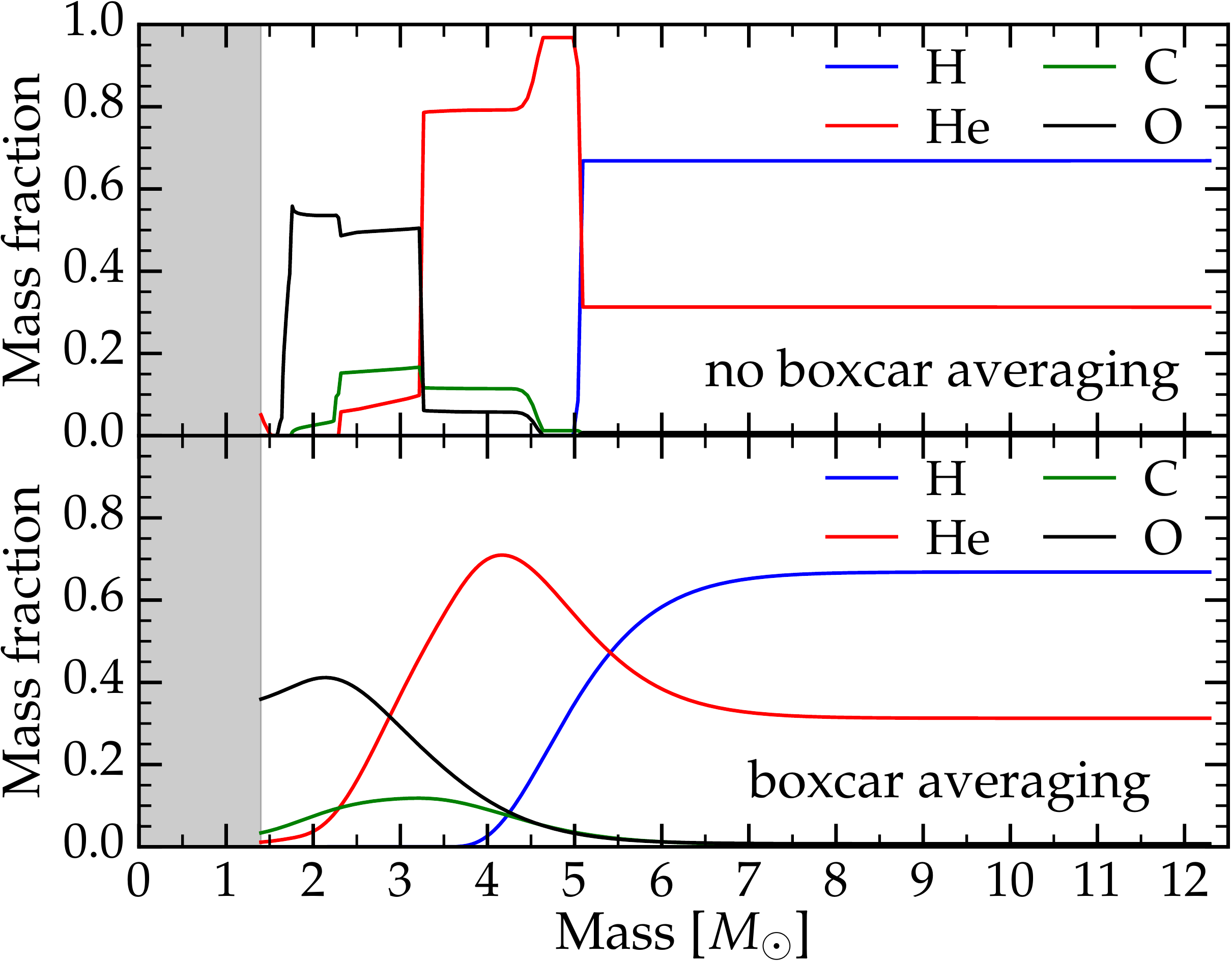}
\caption{Mass fractions of the key elements hydrogen, helium, oxygen,
  and carbon in the unstripped reference model before (upper panel)
  and after (lower panel) boxcar smoothing. Note that the smoothed
  profile here is shown for a grid with uniform spacing in mass. In
  our production setup that uses a non-uniform grid with higher
  resolution in the innermost and outermost regions, small jumps in
  the composition profiles remain, but have no influence on the light
  curve. We excise the mass inside the shaded regions before launching
  the explosion.} \label{fig:composition}
\end{figure}

\section{Light Curves}  
\label{Light_curves}

We next present our study of the light curves from exploding the
stripped \texttt{MESA} models described above. We begin by summarizing
our basic setup in Section \ref{sec:explosion_setup}. We then present
the light curve of an unstripped model in Section \ref{unstripped}
that will serve as a reference for the subsequent discussion of
stripped models. We also include a discussion of how details such as
the mixing, the nickel distribution, and how the explosion is
initiated impact the light curves in Section
\ref{sec:microsens}. Finally, we present our main study of the light
curves from models with varying levels of stripping in Section
\ref{MESA_results}.

\subsection{Explosion and Light Curve Setup}
\label{sec:explosion_setup}

In all explosion models, we excise the inner $1.4\,M_{\odot}$,
assuming that this part collapsed and formed a neutron star. We then
map (via linear interpolation) the hydrodynamic and compositional
variables from \texttt{MESA} to a 1000 cell grid in \texttt{SNEC}. We
choose the grid spacing so that resolution is concentrated in the
interior, where the thermal bomb is placed, and near the surface,
where the photosphere is initially located. In our fiducial resolution
calculations, the innermost cell has a mass of $\Delta m =
6.5\times10^{-3}\, M_\odot$ and the surface cell has a mass of $\Delta
m = 6.5\times10^{-5}\,M_\odot$. The lowest
  resolution in our fiducial setup is $\Delta m = 0.065\,M_\odot$ at
  mass coordinates between $\sim$$2.5$ and $\sim$$5\,M_\odot$
  (at around grid cell 100). The mass of cells
  between the innermost cell and the coarsest cell changes according
  to geometric progression with a fixed ratio
    between two consecutive cells $> 1$. Between the coarsest cell
  and the surface cells we refine by geometric progression with
  a fixed ratio between two consecutive cells
    $< 1$. Examples of the grids may be found
    in Appendix~A.  The release version of \texttt{SNEC} contains a
routine to generate a variety of grid setups.

The explosion is initiated by applying a ``thermal bomb'' to the
innermost region of the model, just above the mass cut in a way
similar to previous work (e.g.,
\citealt{blinnikov:93,bersten:11,aufderheide:91}). The energy of the
bomb is added to the right-hand-side of Equation~(\ref{lag2}) during a
chosen time interval and in a range of mass from the inner boundary,
both exponentially attenuated. For our fiducial model, summarized in
Section \ref{unstripped}, the bomb is spread over $0.02\,M_{\odot}$
and injected in $1\,\mathrm{sec}$.  The thermal bomb mechanism,
implemented this way, typically gives a few percent more energy to the
system than dialed-in, due to the smooth exponential
attenuation. This small excess in energy is recorded and
  accounted for in \texttt{SNEC}'s global energy balance. We find that
  \texttt{SNEC} conserves total energy to better than $1\%$ in a
  full-physics model followed to 150 days past explosion.

An alternative way of phenomenologically modeling an SN explosion is
the ``piston mechanism,'' which has been used, for example, in the
work by \cite{eastman:94}, \citet{utrobin:07}, \citet{kasen:09} and
\citet{dessart:13}. For the same amount of injected energy, the
thermal bomb and piston mechanism give nearly identical light curves,
as was discussed in \cite{bersten:11} and confirmed by our own
simulations. However, when a reaction network is included, piston and
thermal bomb may result in different nucleosynthetic yields
\citep{young:07}. We have implemented a piston in \texttt{SNEC}, but
do not use it in this work, since the thermal bomb makes it easier to
control the energy of the explosion.

It has long been realized in SN light curve modeling that sharp
gradients in the composition profiles of the progenitors may result in
artificial light curve features that are not observed in real SNe. For
example, \citet{utrobin:07} points out a pronounced bump at the end of
the plateau, followed by an abrupt decrease of the bolometric
luminosity for a model with unmixed chemical composition (see their
Figure 16). Two- and three-dimensional simulations of shock
propagation in core-collapse SN explosions (e.g.,
\citealp{kifonidis:03,kifonidis:06,wongwathanarat:15}) show that
effective mixing occurs during the shock propagation through the
progenitor due to the Rayleigh-Taylor and Richtmyer-Meshkov
instabilities.  In this process, hydrogen and helium get mixed into
the inner layers, while metal-rich clumps, and, in particular,
$^{56}\mathrm{Ni}$, may penetrate outwards up to $\sim$$
3000\,\mathrm{km\ s^{-1}}$ and more in the velocity profile. Lacking a
physical mechanism for the mixing in our one-dimensional code, we
apply an artificial ``boxcar'' averaging, as used, for example, in
\citet{kasen:09} and \citet{dessart:12,dessart:13}. We run a boxcar
with a width of $0.4\,M_{\odot}$ through the model four times until we
obtain a smooth profile (details of this procedure are available in
the \texttt{SNEC} notes document on the \texttt{SNEC} website). As an
example, Figure~\ref{fig:composition} depicts the non-mixed (top
panel) and mixed (bottom panel) mass fractions of hydrogen, helium,
oxygen, and carbon in the unstripped reference model.

Finally, we assume a fixed amount of $^{56}\mathrm{Ni}$ of
$0.05\,M_{\odot}$ in all our models, which is roughly the average
amount deduced for SNe IIP (which have a range of $\sim
0.01-0.1\,M_{\odot}$, see \citealp{kasen:09,smartt:09}). We distribute
it uniformly in the interval between the excised mass of
$1.4\,M_{\odot}$ and some chosen mass coordinate ($5\,M_{\odot}$ for
the reference runs, which is near the edge of the helium core) at the
expense of other elements before smoothening the composition.

\begin{figure}
\centering
\includegraphics[width=0.475\textwidth]{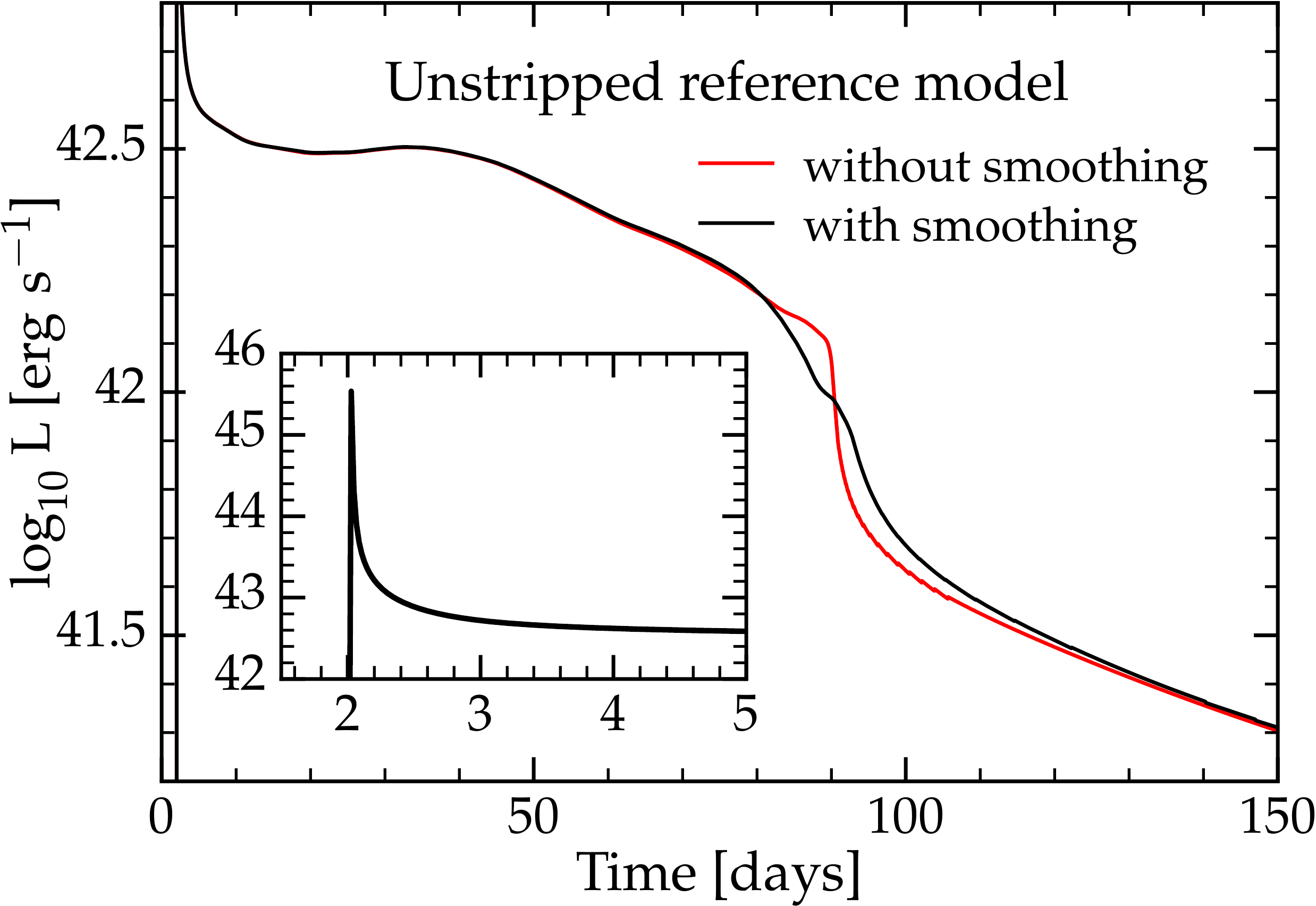}
\caption{Bolometric light curve of the unstripped reference
  model. Time is given relative to the onset of the thermal bomb
  driving the explosion. Shock breakout occurs at day $2.03$ through
  the reference model's surface at $\sim$$1039\,R_\odot$.  The black
  graph is the fiducial light curve obtained with our standard
  parameter choices, including boxcar smoothing as an approximation of
  mixing during the explosion (cf.~Figure~\ref{fig:composition} and
  \S\ref{sec:explosion_setup}). The red graph represents the unmixed
  case with steep compositional gradients (top panel of
  Figure~\ref{fig:composition}). The inset plot shows shock breakout
  and the very early light curve. We note that during shock breakout
  the photosphere is located in the outermost cell of \texttt{SNEC}'s
  grid and spatially poorly resolved. Thus the light curve predicted
  by \texttt{SNEC} around the time of shock breakout is likely
  not reliable (cf.~\citealt{ensman:92}).} \label{fig:lum_smoothed}
\end{figure}

\begin{figure}[t]
\centering
\includegraphics[width=0.47\textwidth]{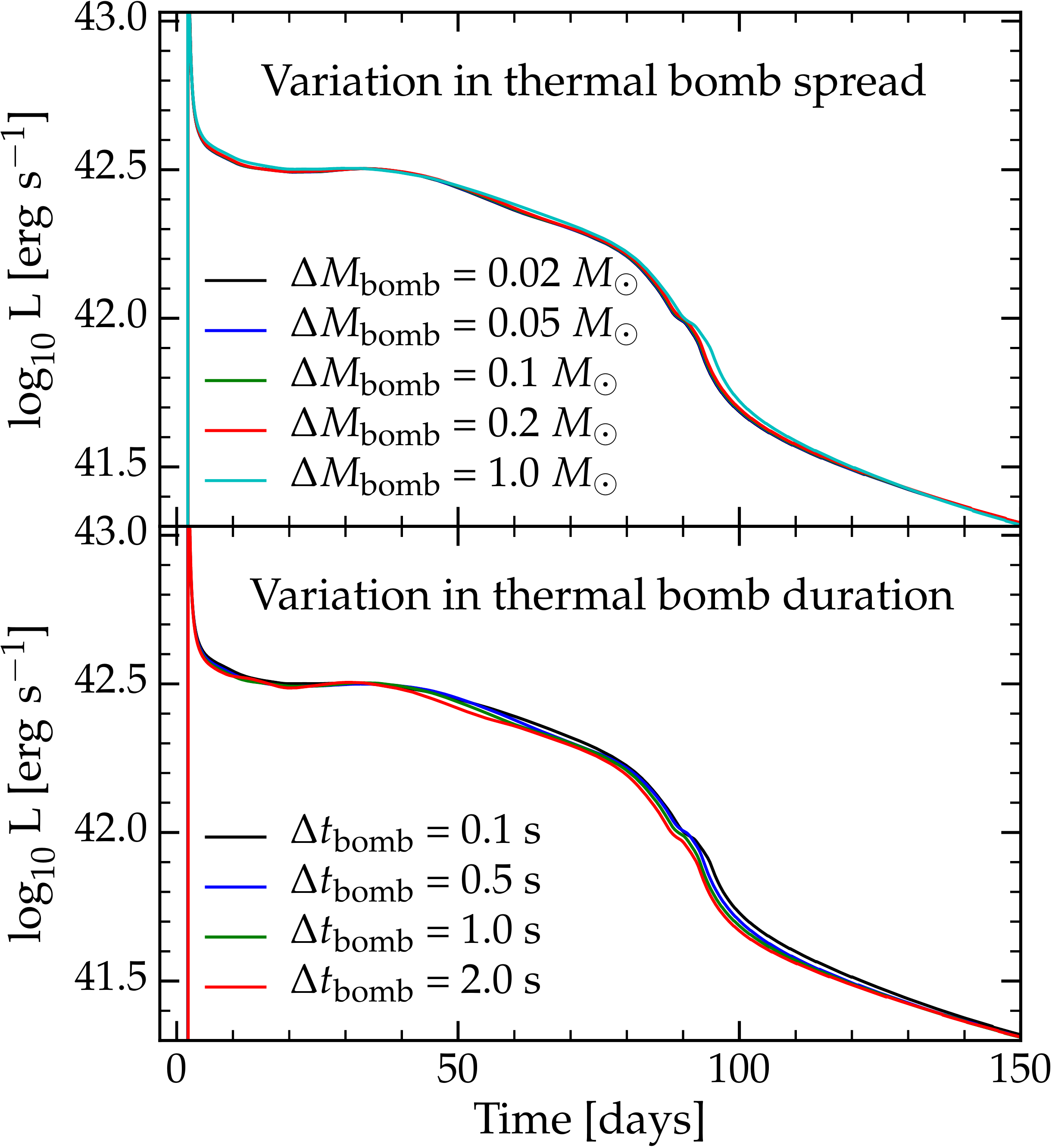}
\caption{Bolometric light curves of the unstripped reference
  progenitor computed with different mass range over which the thermal
  bomb is spread (top panel) and different durations of the thermal
  bomb (bottom panel).  All other parameters are those laid out in
  \S\ref{sec:explosion_setup}.  Time is relative to the onset of the
  thermal bomb. } \label{fig:bomb_parameters}
\end{figure}

\begin{figure}
\centering
\includegraphics[width=0.475\textwidth]{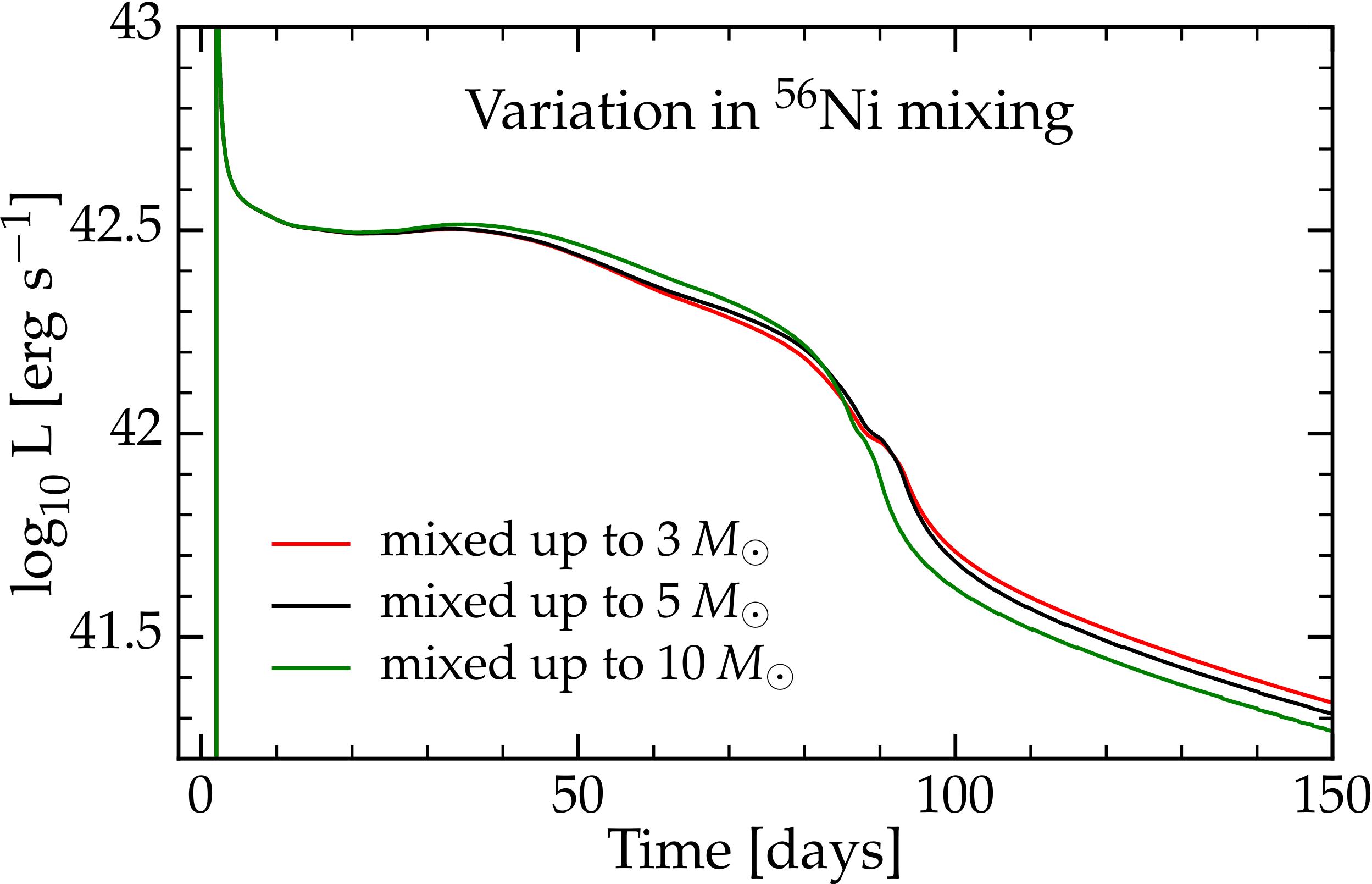}
\caption{Bolometric light curves for the unstripped reference
  progenitor model with different degree and extent of initial
  $^{56}\mathrm{Ni}$ mixing and all other parameters as laid out in
  \S\ref{sec:explosion_setup}. The helium core mass is $5.1\,M_\odot$
  and if $^{56}\mathrm{Ni}$ is mixed smoothly into the hydrogen-rich
  envelope (green graph), then the light curve's ``knee'' feature
  visible when the photosphere drops into the helium core
  disappears.} \label{fig:Ni_mixing}
\end{figure}

\subsection{Fiducial Light Curve of the Unstripped Reference Progenitor Model}
\label{unstripped}

Figure~\ref{fig:lum_smoothed} shows the light curve of the unstripped
reference model. The explosion was initiated at $t = 0$ with a thermal
bomb resulting in an asymptotic kinetic energy of $E =
10^{51}\,\mathrm{erg}$ (as described in
\S\ref{sec:explosion_setup}). We added $M_{\rm Ni} = 0.05\,M_{\odot}$,
mixed up to a mass coordinate of $5\,M_{\odot}$ (but not into the
hydrogen envelope, though boxcar smoothing introduces some nickel into
the hydrogen-rich region.). These are the fiducial explosion parameter
choices used throughout this study.

The light curve of the unstripped reference model shows all the
traditional hallmarks of a typical SNe IIP (e.g.,
\citealt{falk:77,eastman:94,filippenko:97}). Shock breakout occurs
when the optical depth is less than $\sim c/v$ at a time of
$2.03\,{\rm days}$ after the onset of explosion. The bolometric
luminosity peaks at $L=3.4\times
10^{45}\,\mathrm{erg}\,\mathrm{s}^{-1}$ with an effective temperature
of $T_{\rm eff}=1.7\times 10^5\,\mathrm{K}$.  The subsequent cooling
phase (discussed in detail in \citealt{nakar:10}) lasts for
$\sim19\,\mathrm{days}$. At this point, the ejecta have expanded and
cooled so much that hydrogen recombination sets in (starting already
at $T\sim 7500\,\mathrm{K}$) and powers the plateau phase with very
slowly decreasing effective temperature that varies from $\sim
6000\,\mathrm{K}$ at $\sim35\,\mathrm{days}$ down to $ \sim
5000\,\mathrm{K}$ at $\sim90\,\mathrm{days}$. The recombination wave
and, consequently, the photosphere moves inward in mass coordinate,
but due to the overall expansion stays at roughly constant radius,
resulting in a relatively small variation in luminosity during the
 recombination phase of the plateau from day $\sim19$ to
day $\sim80-90$ (this phase is investigated analytically 
in \citealt{goldfriend:14}). The slow decline that is apparent in the plateau
phase shown in Figure~\ref{fig:lum_smoothed} occurs because of the
combined effects of the photosphere receding slightly in radius and
the effective temperature slowly decreasing (e.g.,
\citealt{eastman:94,woosley:88}). 

The plateau ends when the photosphere reaches the helium core. Helium
recombines at $T \gtrsim 10^4\,\mathrm{K}$ whereas the photospheric
temperature is $T \sim 5000\,\mathrm{K}$, recombination accelerates
dramatically, and both the radius of the photosphere and the
luminosity decrease rapidly. Note that it is common in
  the SN IIP theoretical light curve literature to define the
  \emph{plateau duration} as the time from shock breakout to the drop
  when the photosphere reaches the helium core
  \citep{kasen:09,popov:93}. We adopt this definition of plateau
  duration in this paper.  The small ``knee'' or ``bump'' feature in
the drop of the fiducial light curve around day 90 in
Figure~\ref{fig:lum_smoothed} is due to the additional luminosity
input from radioactive $^{56}$Co (from the
$^{56}$Ni$\rightarrow$$^{56}$Co$\rightarrow$$^{56}$Fe decay chain)
that is uncovered as the photosphere sweeps through the helium core
throughout which $^{56}$Ni was mixed initially. This feature is
sensitive to the degree and implementation of mixing and is unlikely
to be robust (see next Section \ref{sec:microsens}). Finally, the tail
of the light curve, after day $\sim$$100$, is powered exclusively by
the radioactive decay of $^{56}$Co.

\subsection{Sensitivity of the Fiducial Light Curve to Mixing,
Thermal Bomb Parameters, and Nickel Distribution}
\label{sec:microsens}

The red curve in Figure~\ref{fig:lum_smoothed} highlights the effect
of steep compositional gradients on the light curve in comparison with
the result obtained with compositional smoothing (black graph;
``boxcar averaging''; Section \ref{sec:explosion_setup}) that we use to
mimic multidimensional mixing during the explosion.  If exploded
without smoothing, hydrogen-rich material transitions discontinuously
to helium-rich material (cf.~Figure~\ref{fig:composition}), which
leads to more rapid recombination, a more abrupt drop of the
photosphere radius, and a steeper decline of the luminosity. Although
observations of most SNe do not generally reveal such abrupt drops,
some subclasses of SNe may have rapidly dropping photospheric
velocities as discussed by \citet{piro:14} because of this same effect
of rapid helium recombination. For all other light curves presented in
this paper, we use the smoothed composition profiles.

In Figure~\ref{fig:bomb_parameters}, we explore the sensitivity of the
fiducial light curve to (top panel) variations in the amount of mass
over which the thermal bomb is spread and (bottom panel) variations in
the duration over which the energy is injected. While the details of
the energy injection will depend on the actual physical explosion
mechanism (e.g., \citealt{bethe:90,janka:12a}), it is reassuring that
the light curve is fairly insensitive to both mass spread and duration
of energy injection. We remind the reader that for the light curves shown in Sections
    \ref{unstripped} and \ref{MESA_results}, the thermal bomb is spread
    over $0.02\,M_{\odot}$ and its duration is $1\,\mathrm{s}$ (in
    self-consistent multi-dimensional core-collapse SN simulations 
    most of the energy injection appears to occur within
    $\sim1-2\,{\rm s}$; e.g., \citealt{bruenn:13}).

Finally, Figure~\ref{fig:Ni_mixing} shows the dependence on
$^{56}\mathrm{Ni}$ mixing. The overall effect of $^{56}\mathrm{Ni}$
mixing is modest.  The general trend is that with the initial mixing
of $^{56}\mathrm{Ni}$ to increasing mass coordinates, the contribution
to the luminosity of the
$^{56}$Ni$\rightarrow$$^{56}$Co$\rightarrow$$^{56}$Fe decay chain
becomes more prominent. At later times, the luminosity due to
radioactive decay becomes smaller with increasing mixing due to the
fact that more gamma rays escape from the model without being absorbed
\citep[as also described by][]{young:04,utrobin:07,bersten:11}. Note
that if $^{56}\mathrm{Ni}$ is mixed far into the hydrogen-rich
envelope, the ``knee'' feature at the end of the plateau
disappears. This is consistent with the findings of \cite{kasen:09},
who mixed $^{56}\mathrm{Ni}$ into the hydrogen-rich envelope in their
models.

\subsection{Light Curves as a Function of Mass Stripping}
\label{MESA_results}

\begin{figure}
\centering
\includegraphics[width=0.475\textwidth]{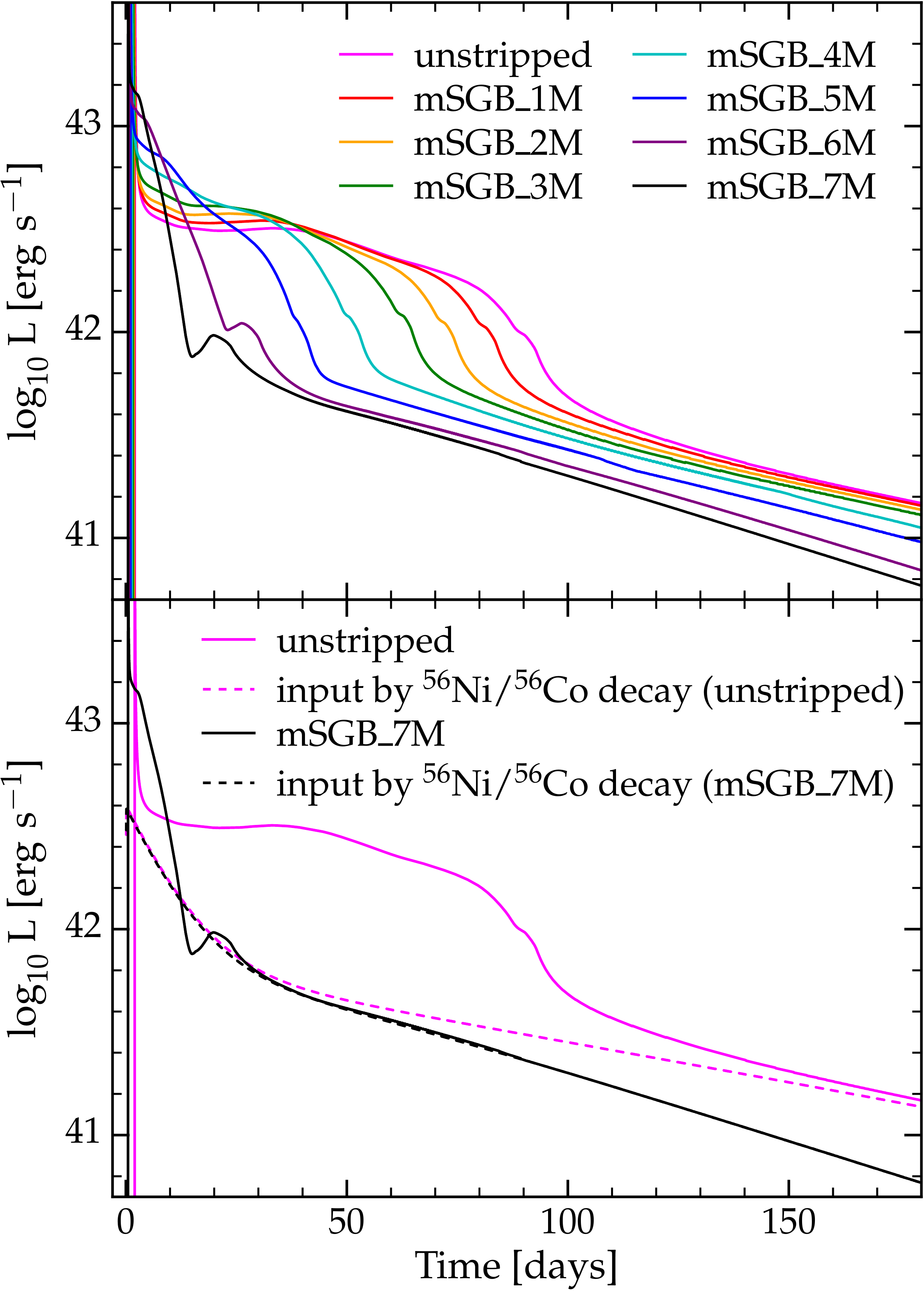}
\caption{Upper panel: Bolometric light curves from the mSGB grid of
  progenitor models.  Lower panel: Light curves from the models
  `unstripped' and `mSGB\_7M'.  Dashed lines show the heating rate
  from the radioactive decay of $^{56}\mathrm{Ni}$ deposited in each
  model after taking into account leakage of the gamma rays. In the
  most stripped model, gamma rays leak out faster than in the
  unstripped model, impacting the late time light
  curve.} \label{fig:lum_full}
\end{figure}

\begin{figure}
\centering
\includegraphics[width=0.485\textwidth]{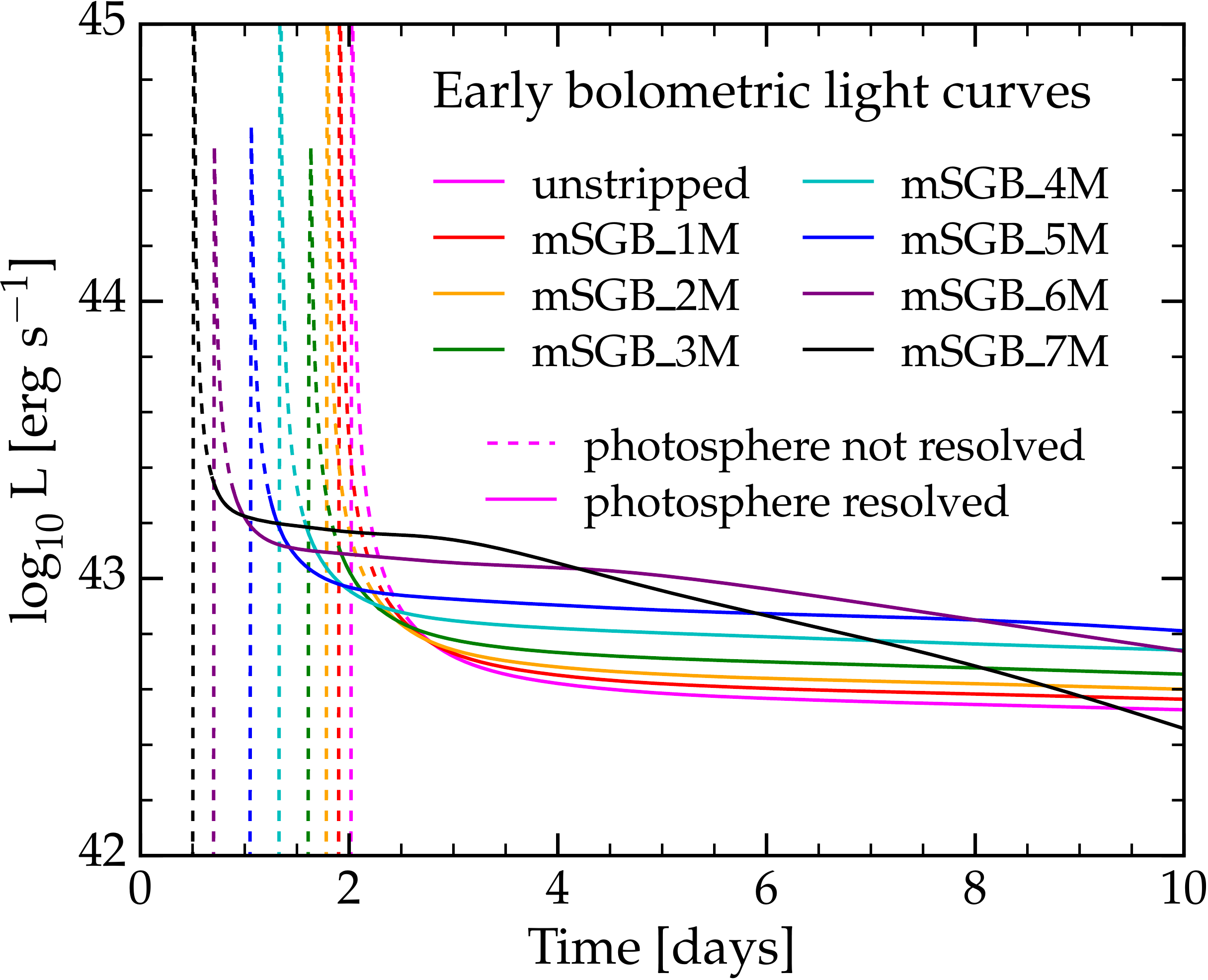}
\caption{The same bolometric light curves as in
  Figure~\ref{fig:lum_full}, but instead focusing on early times
  around shock breakout. Time is given relative to the onset of the
  thermal bomb driving the explosion.  Solid parts of the curves start
  from the time at which the photosphere moves inwards in the grid
  space, while dashed parts indicate that the photosphere is located
  in the outermost grid cell and is spatially poorly resolved. Note
  that more stripped models have a higher luminosity in the
  post-breakout cooling phase and a faster evolving
  light curve.} \label{fig:lum_early}
\end{figure}

Figure~\ref{fig:rho_M} demonstrates that the structure of the
hydrogen-rich envelope and of the outer helium core are essentially
independent of the point at which we remove mass for the set of
stripping points we choose in this study
(cf.~\S\ref{sec:presnstructure}). This suggests that the resulting light
curves should be independent of the stripping point and we check this
assertion later in this section. Here, we operate under the assumption
that it is true and focus our discussion on the model series stripped
at the middle of the subgiant branch (mSGB).

\begin{figure}[t]
\centering
\includegraphics[width=0.485\textwidth]{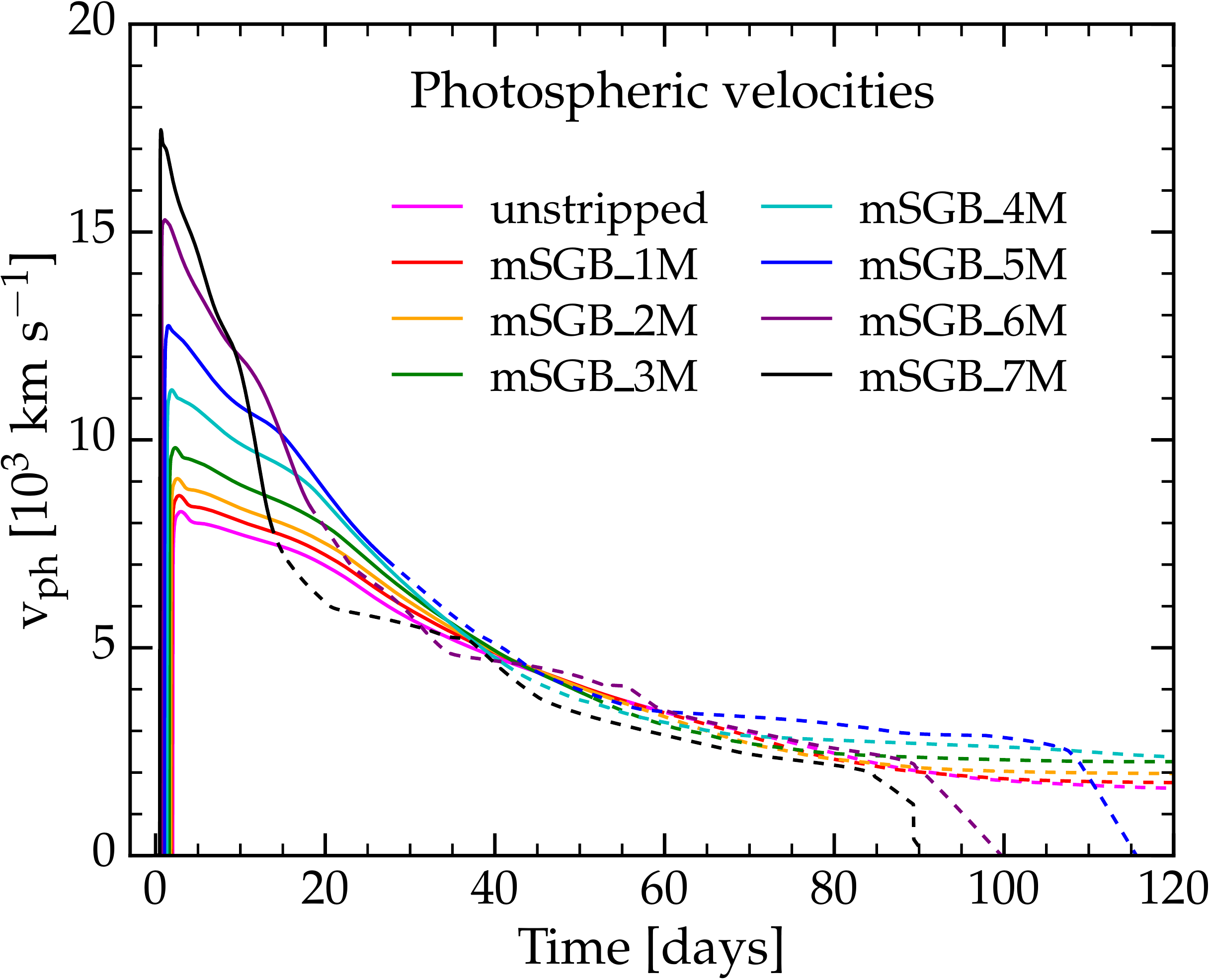}
\caption{Photospheric velocities $v_\mathrm{ph}$ in the mSGB model
  series, including the unstripped reference model. These velocities
  are reliable only until the photospheric temperature drops below $T
  \sim 10^{3.75}\,\mathrm{K}$ below which we cannot accurately estimate
  the location of the photosphere due to low-$T$ opacity limitations
  in \texttt{SNEC}. We indicate the unreliable part of $v_\mathrm{ph}$
  by plotting it in dashed lines. Note that the temperature drops most
  rapidly in the most stripped models that lack a plateau phase
  (cf.\ Figure~\ref{fig:lum_full}).
 \label{fig:velocities}}
\end{figure}

In the top panel of Figure~\ref{fig:lum_full}, we show bolometric mSGB
series light curves obtained for a final ejecta kinetic energy of 
$E_{\rm kin}=10^{51}\,\mathrm{erg}$ and $M_{\rm Ni} = 0.05\,M_{\odot}$
(and all other explosion parameters as specified in
\S\ref{sec:explosion_setup}). The early, shock breakout and cooling
part of the light curves is shown in Figure~\ref{fig:lum_early}.
Shock breakout itself is not well resolved (i.e., the photosphere is in the
outermost grid cell) and thus the light curves in this phase are
unreliable (cf.~\citealt{ensman:92}). Once the photosphere begins to
move inward into the expanding envelope, the light curves predicted by
\texttt{SNEC} become robust. Models with greater amounts of mass
stripped have higher luminosities in the cooling phase and decay more
rapidly. In more stripped models, the SN shock has to propagate
through less envelope mass that is more tenuously distributed. This
results in a higher shock velocity at early times as shown in 
Figure~\ref{fig:velocities} (and earlier
breakout; as shown in Figure~\ref{fig:lum_early}), which leads to a
hotter photosphere and a more rapid expansion of the ejecta. This
translates directly to a higher initial luminosity and a more rapid
decay of the light curve.

From the top panel of Figure~\ref{fig:lum_full} we see that as long as
there is a substantial amount of hydrogen-rich material left, a clear
(if very short) plateau due to recombination can be made out. In our
mSGB model series, this is until models mSGB\_4M and mSGB\_5M, from
which we strip $4\,M_\odot$ and $5\,M_\odot$ and which have
$2.67\,M_\odot$ and $1.61\,M_\odot$ of hydrogen-rich material left,
respectively. The two most stripped models of this series, models
mSGB\_6M and mSGB\_7M, have only $0.74\,M_\odot$ and $0.38\,M_\odot$
of very tenuous hydrogen-rich envelope left, respectively. This does
not appear to be sufficient to lead to any plateau and the photosphere
recedes very quickly in these models. Instead, models
mSGB\_6M and mSGB\_7M show a clear peak between $\sim20-30\,{\rm days}$ that
is analogous to the nickel-powered peak that is seen in all types of hydrogen-deficient (i.e., Type I) SNe.

Figure~\ref{fig:lum_full} shows that stripping of hydrogen-rich
envelope mass also has an effect on the late-time radioactively
powered part of the light curve. The late-time  
light curves exhibit changes with mass stripping because there is earlier
leakage of gamma rays from the more highly stripped models. This is
shown by the lower panel of Figure~\ref{fig:lum_full}. It shows the
unstripped reference model and the most stripped model
(mSGB\_7$M_\odot$) in comparison with the total amount of heating
deposited in each model due to the radioactive
$^{56}$Ni$\rightarrow$$^{56}$Co$\rightarrow$$^{56}$Fe decay chain.
The most stripped model has less radioactive heating at late times.

\begin{figure}[t]
  \centering
  \includegraphics[width=0.48\textwidth]{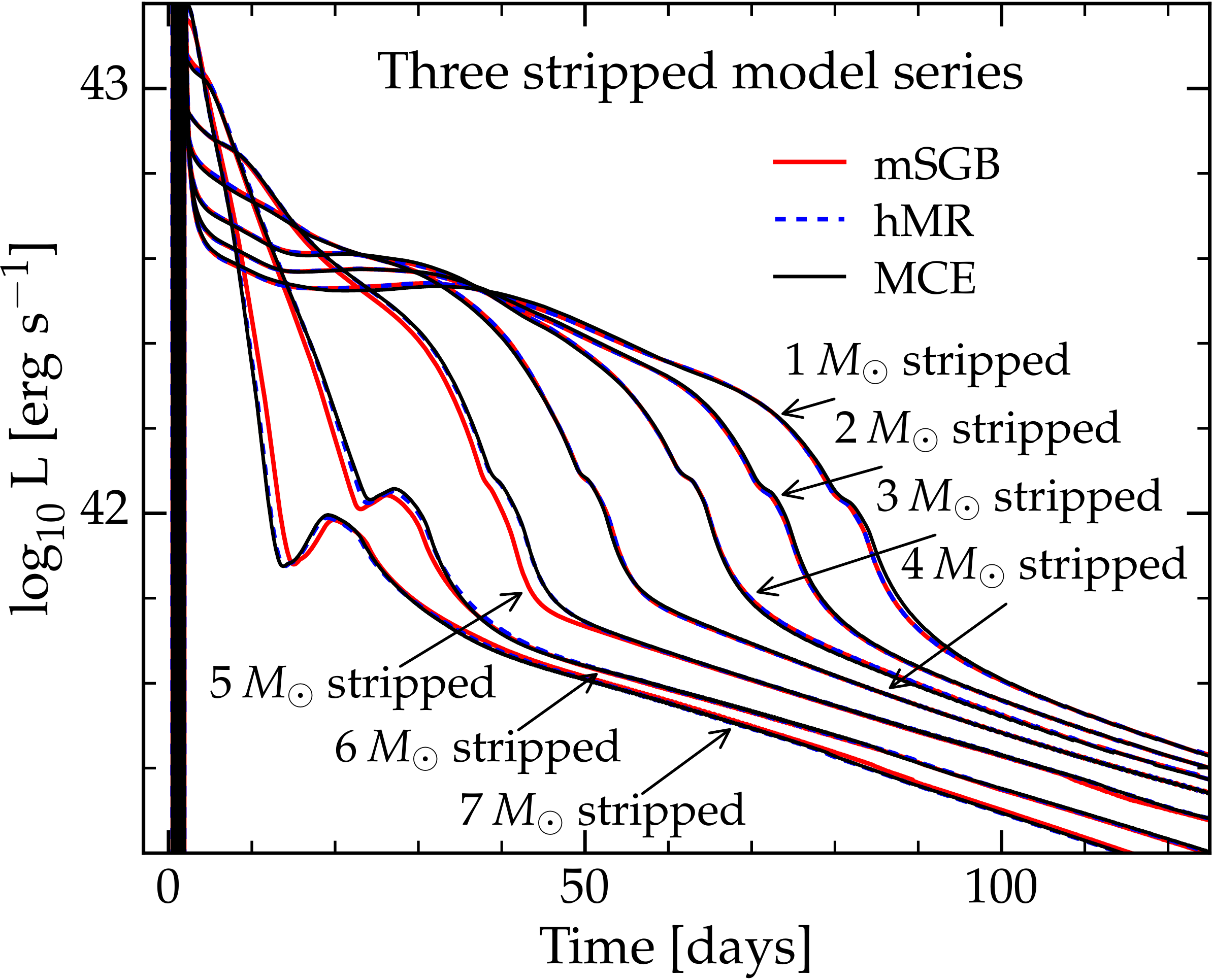}
  \caption{Bolometric light curves for all models from the three
    different \texttt{MESA} series, described in the
    Tables~\ref{tab:stripped_setup} and \ref{tab:stripped_res}. The
    explosion setup is identical in all cases and as described in
    \S\ref{sec:explosion_setup}. All models are stripped after they
    have left the main sequence, but the precise point of stripping
    has little influence on the resulting light
    curve.} \label{fig:lum_MESA}
\end{figure}

Figure~\ref{fig:lum_MESA} compares the bolometric light curves of
models stripped at the middle of the subgiant branch (mSGB, the model
series we focus on), half-maximum radius (hMR), and maximum radial
extent of the convective envelope (MCE); cf.~\S\ref{sec:whenstrip} and
Table~\ref{tab:stripped_setup}. The light curves of models with the
same amount of mass stripped at different times are nearly identical,
supporting our initial assertion on the basis of
Figure~\ref{fig:rho_M}. Any variation between light curves of models
stripped at mSGB, hMR, and MCE is arguably smaller than the level of
systematic uncertainty inherent to \texttt{SNEC}'s approximate (i.e.,
equilibrium diffusion) way of predicting these light curves. The robustness
of the light curves suggests that comparisons with observations may
allow reliable conclusions about the amount of hydrogen left after
(rapid) mass loss events. However, we note that we exclusively consider
post-MS mass loss. Early large-scale mass loss events on the MS may
lead to different outcomes, which should be explored in future work.

\begin{figure*}[t]
\begin{center}
\includegraphics[width=1.0\textwidth]{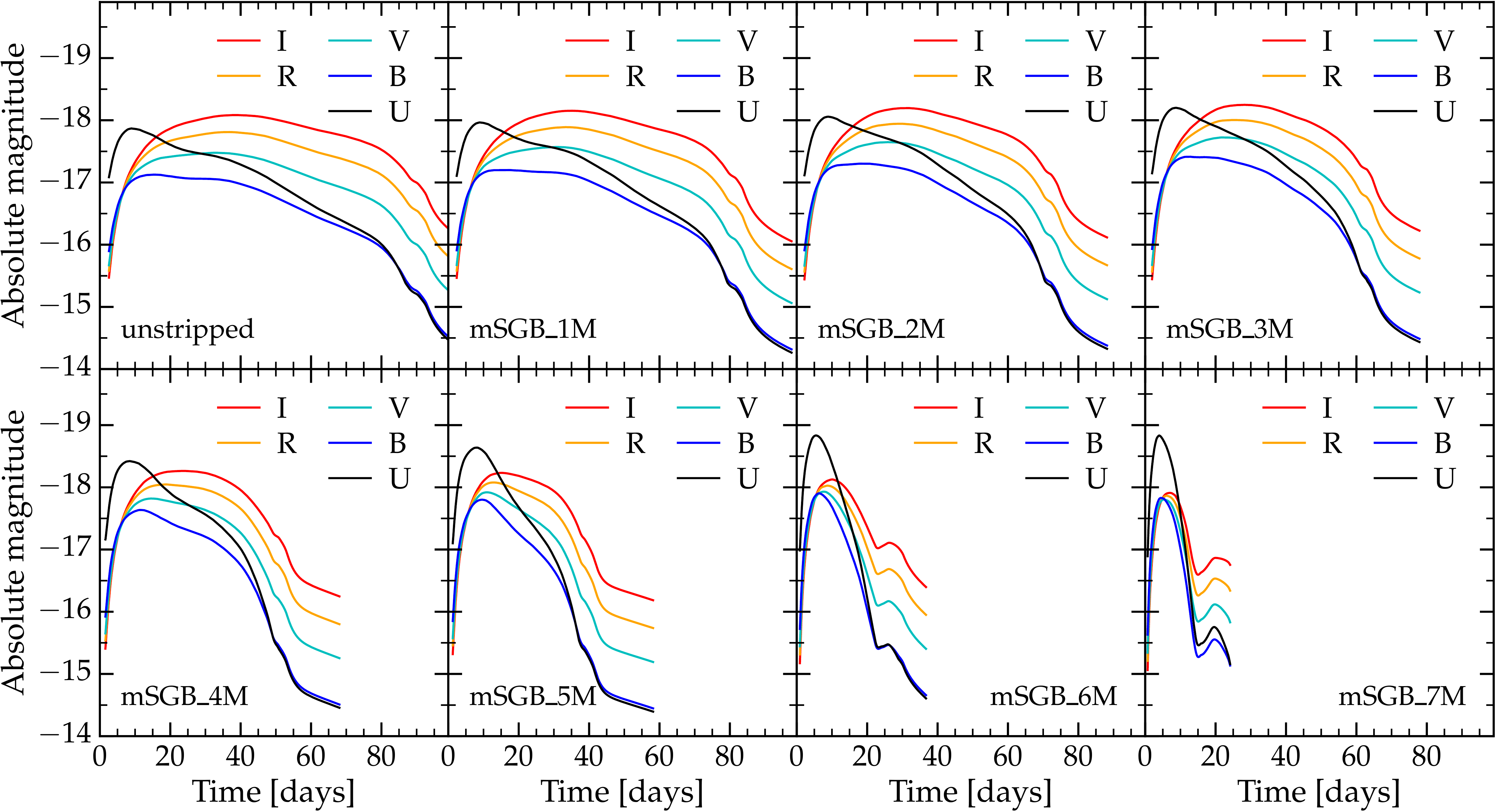}
\end{center}
\caption{Light curves in absolute magnitude in \emph{IRVB} and
  \emph{U} bands obtained with \texttt{SNEC} for our mSGB model set.
  The time is given relative to the onset of energy injection by the
  thermal bomb. The light curves start when the photosphere no longer
  coincides with the outermost grid cell in the \text{SNEC}
  calculation.  Shock breakout, which is in the UV, would be visible
  in \emph{U} band, but is not shown.  The curves are terminated at
  the point at which the explosion begins to transition to the nebular
  phase and the black body approximation underlying the band light
  curves is no longer valid (we define this point at the time at which
  5\% of the luminosity comes from above the photosphere due to gamma-ray
  deposition). We note that real SN light curves fade away in
  \emph{U} and \emph{B} bands considerably faster than predicted by
  \texttt{SNEC} (e.g., \citealt{kasen:09,dessart:13}).  This is due to
  line blanketing by iron group elements that is unaccounted for in
  our models.} \label{fig:mag_IRVBU}
\end{figure*}

\begin{figure}[h]
\centering
\includegraphics[width=0.48\textwidth]{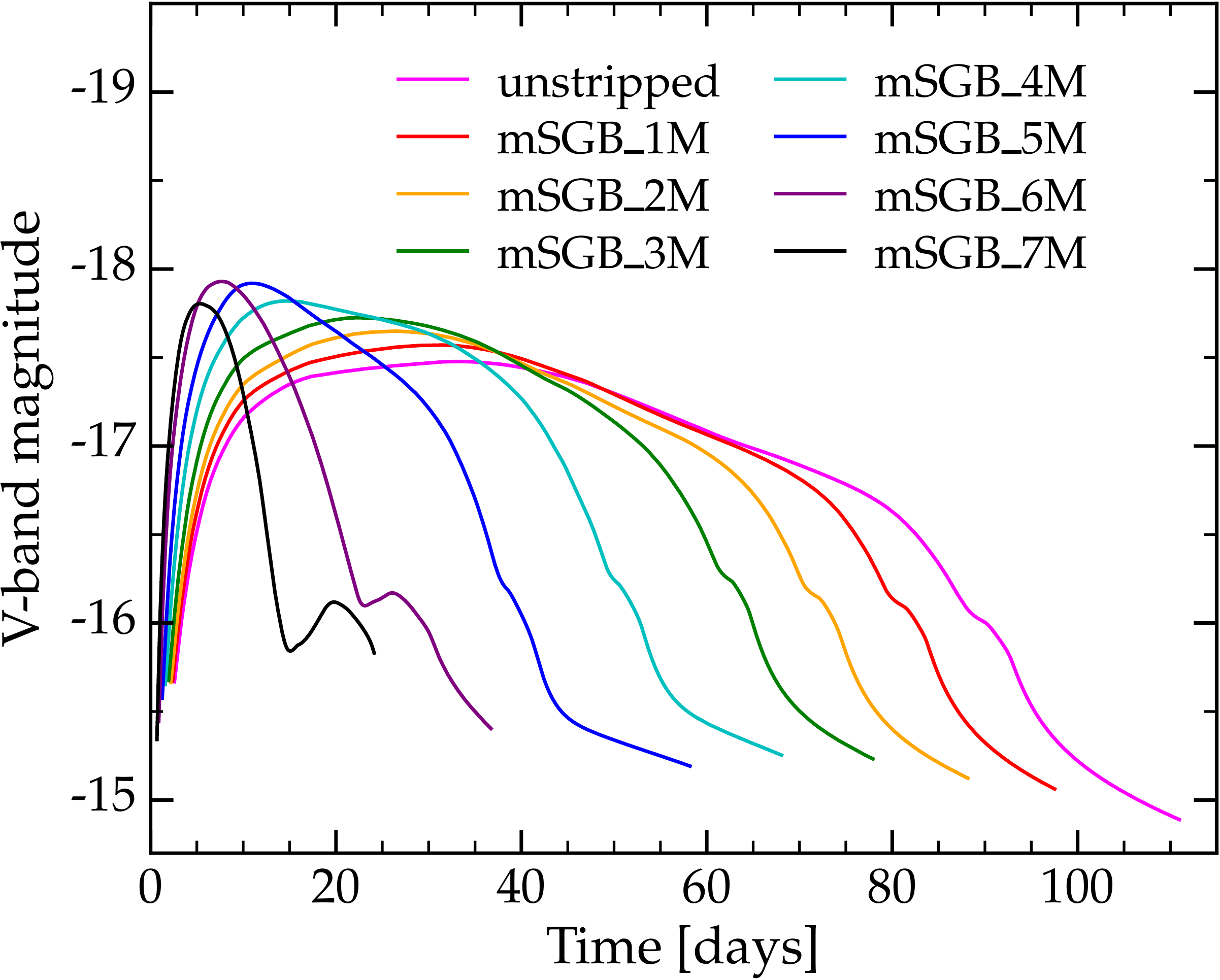}
\caption{Light curves in absolute \emph{V}-band magnitude obtained
  with \texttt{SNEC} for the mSGB model set. The time is given
  relative to the onset of energy injection by the thermal bomb. The
  \emph{V}-band light curves of more stripped models evolve (rise,
  decay) faster than those of less stripped
  models.} \label{fig:mag_V_all}
\end{figure}

In Figure~\ref{fig:mag_IRVBU}, we plot approximate absolute magnitudes
of the mSGB series light curves in the \emph{IRVB}- and \emph{U}-bands
and in Figure~\ref{fig:mag_V_all}, we focus on the \emph{V}-band.  We
obtain the band light curves by assuming black body emission from the
photosphere and using the bolometric corrections from
\cite{ofek:14}. When interpreting these light curves, one should keep
in mind two important caveats: ({\bf 1}) When the whole ejecta becomes
optically thin, the luminosity has a large contribution of
$^{56}\mathrm{Ni}/^{56}\mathrm{Co}$ from above the photosphere. For
this reason, we terminate the curves at the points where the
luminosity contribution due to $^{56}\mathrm{Ni}/^{56}\mathrm{Co}$
above the photosphere amounts to more than $5\%$ of the total
luminosity. ({\bf 2}) As was demonstrated in \citet{kasen:09}, the
\emph{U}- and \emph{B}-bands of the light curves cannot be adequately
reproduced by a one-temperature equilibrium-diffusion code like
\texttt{SNEC}, because these bands are strongly influenced by iron
group line blanketing after a few tens of days (see, e.g., Figure~8 of
\citealp{kasen:09}). This causes a much faster decline of the
\emph{U}- and \emph{B}-band light curves. However, the \emph{IR}- and
\emph{V}-bands are still similar to a single temperature black body
spectrum, and thus these bands are more accurately captured by
\texttt{SNEC}.

\begin{figure}
\centering
\includegraphics[width=0.475\textwidth]{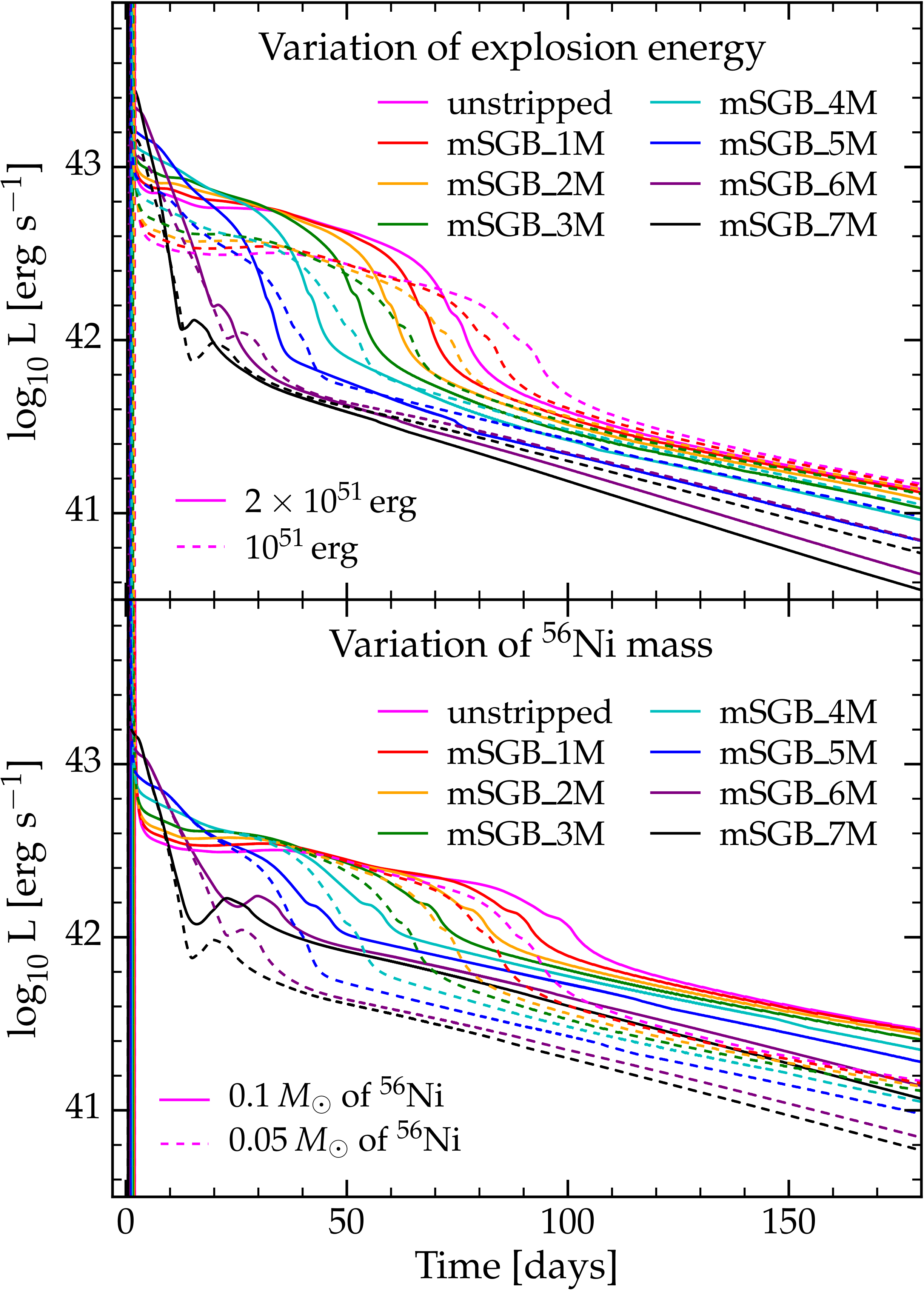}
\caption{Bolometric light curves for the mSGB model set as in
  Figure~\ref{fig:lum_full}, but with variations in explosion energy
  and $^{56}$Ni mass. The time is given relative to the onset of
  energy injection by the thermal bomb. Top panel: Comparison of light
  curves of models with the fiducial final kinetic energy
  ($10^{51}\,\mathrm{erg}$, dashed curves) and with twice that energy
  ($2\times 10^{51}\,\mathrm{erg}$, solid curves). Increasing the
  explosion energy leads to brighter, more rapidly evolving explosions
  in agreement with previous work.
  Bottom panel: Comparison of light curves of models with the fiducial
  $^{56}$Ni mass ($0.05\,M_\odot$, dashed curves) and with twice that
  amount of $^{56}$Ni ($0.1\,M_\odot$, solid curves). More nickel leads
  to extended plateaus and brighter radioactive tails.
  The qualitative changes due to variations in explosion energy and
  $^{56}$Ni mass are in agreement with what was found in previous work
  (e.g., \citealt{kasen:09,young:04}).} \label{fig:lum_double_energy}
\end{figure}

Finally, we study the sensitivity of our light curves to doubling the
explosion energy to $2\times 10^{51}\,\mathrm{erg}$ and to doubling
the amount of initially present $^{56}$Ni to $0.1\,M_\odot$ in
Figure~\ref{fig:lum_double_energy}. The qualitative light curve
changes are overall as expected from previous work (e.g.,
\citealt{young:04,utrobin:07,kasen:09}): More energetic explosions
have brighter, but faster evolving light curves and an increased
amount of $^{56}$Ni prolongs the plateau and results in a higher
late-time luminosity. Increasing the amount of $^{56}$Ni also results
in a more pronounced second light curve peak in the most-stripped
models mSGB\_6M and mSGB\_7M. We summarize the connection between
these calculations and observed SN light curves in the discussion
below.

\section{Discussion and Conclusions}
\label{Conclusions}

We presented the new open-source SuperNova Explosion Code
(\texttt{SNEC}) for investigating supernova (SN) explosions and the
resulting light curves. \texttt{SNEC}, while currently limited to
equilibrium (single temperature) radiative diffusion, is the first
such code that is publicly available and this paper will serve as a
reference and starting point as \texttt{SNEC} is utilized and improved
by both us and the broader community in the future.

As a first application of \texttt{SNEC}, we studied the explosions of
$M_\mathrm{ZAMS} = 15\,M_\odot$ stars with varying levels of rapid
post-MS mass loss at different times in the stars' evolution from the
main sequence to the supergiant stage. We evolved these stars to the
onset of core collapse with the open-source \texttt{MESA} stellar
evolution code and then exploded them with \texttt{SNEC} using a
thermal bomb resulting in an asymptotic explosion energy of
$10^{51}\,\mathrm{erg}$. At three different times during the evolution
to the supergiant stage, we systematically stripped hydrogen-rich
material in units of $1\,M_\odot$, leaving in the most extreme case
only a thin radiative hydrogen-rich layer above the hydrogen shell
burning zone. In this experiment in massive star evolution, we find
that the time of stripping has essentially no influence on the
structure of the envelope and thus most of the SN light curve. Stars
with more than $1.5-2\,M_\odot$ of hydrogen-rich material left die as
red supergiants with $R \gtrsim 900\,R_\odot$ and our most stripped
star ($0.31\,M_\odot$ of hydrogen-rich envelope left) dies as a yellow
supergiant with a still extended, very tenuous envelope of $R \sim
550\,R_\odot$.

We find that the time of stripping and the amount of mass stripped has a
big but not clearly systematic effect on the structure of the layers
immediately surrounding the iron core (mass coordinate
$\sim$$1.5-3.3\,M_\odot$). The structure in this particular region has
been shown to be highly relevant for deciding the ultimate outcome of
core collapse (explosion/no explosion, black hole/neutron star
remnant; \citealt{oconnor:11,ugliano:12,ertl:15}). Our results suggest
the need for a detailed study of the sensitivity of presupernova
stellar structure to large amounts of rapid mass loss (e.g., via
unstable mass transfer in a binary).

The light curves resulting from our set of presupernova models show SN
IIP-like morphology for models with more than $\sim$$1.5-2\,M_\odot$ of
hydrogen-rich envelope material left at the presupernova stage. The
most stripped models ($\lesssim$$1.5\,M_\odot$ of hydrogen-rich envelope
left) have higher luminosity in the post-breakout cooling phase, but
show no plateau, but a second peak around $20-30\,{\rm days}$ due to
energy input from the radioactive decay of $^{56}$Ni/$^{56}$Co that is
uncovered by the rapidly receding photosphere in these models.

In those models that show a plateau in their bolometric
  light curves, the duration of the plateau phase varies in the
range $\sim20-100\,\mathrm{days}$ (we include both the
  cooling and the recombination phases of the plateau in the plateau
  duration; \citealt{kasen:09,popov:93}), with plateau length
decreasing with decreasing mass of hydrogen-rich envelope
material. In nature, most SNe IIP show plateaus of $\sim
  80-100\,{\rm days}$ \citep[e.g.,][]{poznanski:09,arcavi:12},
  although there may be some evidence for a subset of shorter
  plateaus \citep{anderson:14}. The
   completely mass stripped SNe Ib/c show no plateau at all.
  Both SNe~IIP and SNe Ib/c are relevant
to our study given the inference that many SN IIP progenitors have
ZAMS masses around $\sim15\,M_\odot$ \citep{smartt:09} and arguments
that most SNe Ib/c must come from a similar mass range
\citep{smith:11}. The apparent paucity of {\it observed} short and
  intermediate-length plateaus suggests that in nature hydrogen mass
loss is an all or nothing process, at least for the 
 ZAMS mass we consider. This is perhaps not surprising given that for $M_{\rm
  ZAMS}\lesssim20\,M_\odot$ radiative driven winds are rather weak in
normal prescriptions and appreciable mass loss can probably only occur
from events like binary interactions, which would not be expected to,
say for example, rip off just $\sim50\%$ of the mass.

Our most stripped models still have $\approx0.3-0.4\,M_\odot$ of hydrogen
present and thus should make some connection with SNe IIb. Indeed, the
light curves of these progenitors show two distinct peaks, similar to
the morphology of many SN~IIb, where the first peak comes from the
shock cooling of the remaining surface hydrogen and the second from
radioactive heating \citep{woosley:94,bersten:12,nakar:14}. However,
in detail, the
width of the first peak is too large, which is likely due to our models
having too much hydrogen still present \citep{nakar:14}.  In addition,
the second peak in our light curves is sometimes too dim in comparison to
observed SNe IIb due to our models having somewhat less $^{56}$Ni \citep{lyman:14}. However,
there is a lot of diversity and uncertainty in the amount of $^{56}$Ni
produced in SNe IIb
\citep[see the work and discussions in][]{shigeyama:94,bersten:12,ergon:15},
thus we plan to investigate this in more detail in future work.

Another interesting connection to consider is how our results relate
to SNe IIL. There has been a long-standing
discussion on whether they form a continuous sequence of events that
smoothly transition to SNe~IIP.  The idea that SNe~IIL are instead
distinct from \mbox{SNe IIP} has been argued for by \citet{arcavi:12}
and \citet{faran:14b}, but more recently it has been shown that
SNe~IIL actually show a significant drop in their light curves at late
times \citep[$\sim$$100\,{\rm days}$,][]{valenti:15}, much like SNe
IIP.  If there was a continuous range of events from SNe~IIL to
SNe~IIP, then a natural physical mechanism to consider is gradual loss
of the outer hydrogen, where SNe~IIL would be on the hydrogen-poor
side. At least for the ZAMS mass we consider here 
($15\,M_\odot$), this does not
appear to be the case. First, our results show that intermediate
levels of hydrogen mass loss simply shorten the plateau length which
is different from SNe~IIL, which appear to have roughly normal
duration, but steeply declining ``plateaus'' when they
are followed for a sufficient amount of time
\citep{valenti:15}. Second, SNe~IIL are on average more luminous than
SNe~IIP by $\sim1.5\,{\rm mag}$ in the optical during the first
$\sim10\,{\rm days}$
\citep{patat:93,patat:94,anderson:14,faran:14b,sanders:15}.  Our more
stripped models show slightly higher luminosities at early times, but
not nearly extreme enough. Overall, however, our findings, in
combinations with recent observations \citep{valenti:15}, appear to
argue that perhaps SNe IIL do not necessarily have less hydrogen, but
the hydrogen mass is distributed in a different way. The brighter
early light curves would argue that SNe~IIL have material at a larger
radius \citep{piro:13}. The occurrence of narrow line features in SNe
IIn that might otherwise look somewhat like an SN~IIL \citep{smith:15}
might argue for some contribution from circumstellar
material. \texttt{SNEC} is well-suited for addressing these ideas in a
systematic way in future work since various mass and density
distributions can easily be implemented to investigate what is in fact
needed to reproduce SN~IIL light curves.

Future work will be directed toward exploiting \texttt{SNEC}'s current
capabilities for the systematic and \emph{reproducible} light curve
modeling for a broad range of SN explosions, but also toward improving
\texttt{SNEC}'s transport solver and opacity microphysics. In a first
step, we will upgrade \texttt{SNEC} to handle separate radiation and
matter temperatures with the long-term goal of constructing an
open-source multi-group radiation-hydrodynamics code. Of course, input
from the community will be especially critical for steering \texttt{SNEC}'s
further evolution and we look forward to the community's feedback.

\acknowledgments We acknowledge helpful discussions with and feedback from
J.~P.~Anderson, W.~D.~Arnett,
M.~Bersten, A. Burrows, L.~Dessart, C.~Fryer, M.~Modjaz,
E.~S.~Phinney, D.~Radice, S.~N.~Shore, N.~Smith, A.~Soderberg, and
C.~Wheeler. We thank M.~Bersten for providing us with the initial
conditions necessary to reproduce her results on SN 1999em and also
for generously answering our questions. We thank L.~Dessart for
helping us better understand the discrepancies between \texttt{SNEC}
light curves and his results. We thank B.~W.~Mulligan for helping us
to develope the code. Some of the ideas underlying this study
were inspired by discussions at Palomar Transient Factory Theory
Network workshops at the Sky House, Los Osos, CA. \texttt{SNEC} is
available as open source from
\url{http://stellarcollapse.org/SNEC}. We thank Frank Timmes for
allowing us to use and distribute with \texttt{SNEC} his equation of
state and Saha solver routines. We thank the OPAL opacity project, in
particular Carlos Iglesisas, allowing us to distribute their
interpolation routines and opacity tables with \texttt{SNEC}.  We
thank Jason Ferguson for allowing us to distribute the low-temperature
opacity tables of \cite{ferguson:05} with \texttt{SNEC}. This work is
supported in part by the National Science Foundation under award
Nos.\ AST-1205732 and AST-1212170, by Caltech, and by the Sherman
Fairchild Foundation. The computations were performed on the Caltech
compute cluster Zwicky (NSF MRI-R2 award no.\ PHY-0960291), on the NSF
XSEDE network under allocation TG-PHY100033, and on NSF/NCSA Blue
Waters under NSF PRAC award no.\ ACI-1440083.

\bibliographystyle{apj}
\bibliography{SNEC_manuscript}


\renewcommand \theequation {A.\arabic {equation}}
\setcounter{equation}{0}

\appendix
\section{Comparison with Bersten~et~al.\ (2011)}
\label{Ap1}

The work of \citet{bersten:11} is among the codes that are most
similar to \texttt{SNEC}, so it is important to compare light
curves. This is done using their progenitor model for SN 1999em,
kindly provided by the authors (details of the structure and
composition of the model may be found in their work). We use the same
explosion energy of $E=1.25\times10^{51}\,\mathrm{erg}$.
We also use a step function for the opacity floor,
  i.e., $0.01\,\mathrm{cm}^2\,\mathrm{g}^{-1}$ for material with $Z\le
  0.3$ and $0.24\,\mathrm{cm}^2\,\mathrm{g}^{-1}$ for material with
  $Z>0.3$. This is the only model in this paper where we use this
  opacity law, since we want to explicitly follow \citet{bersten:11}.

\begin{figure}
\begin{center}
\includegraphics[width=0.48\textwidth]{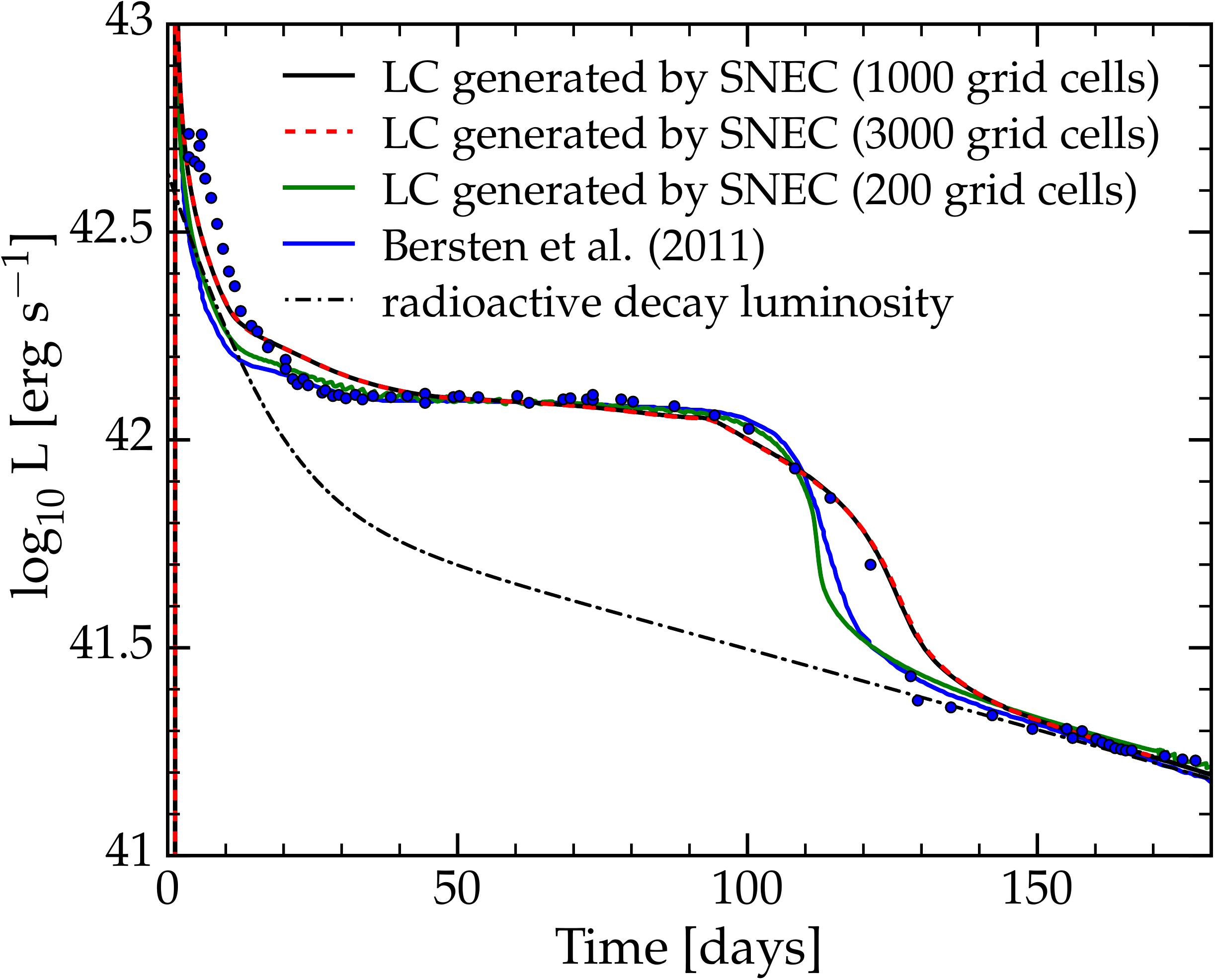}
\hspace{1pt}
\includegraphics[width=0.48\textwidth]{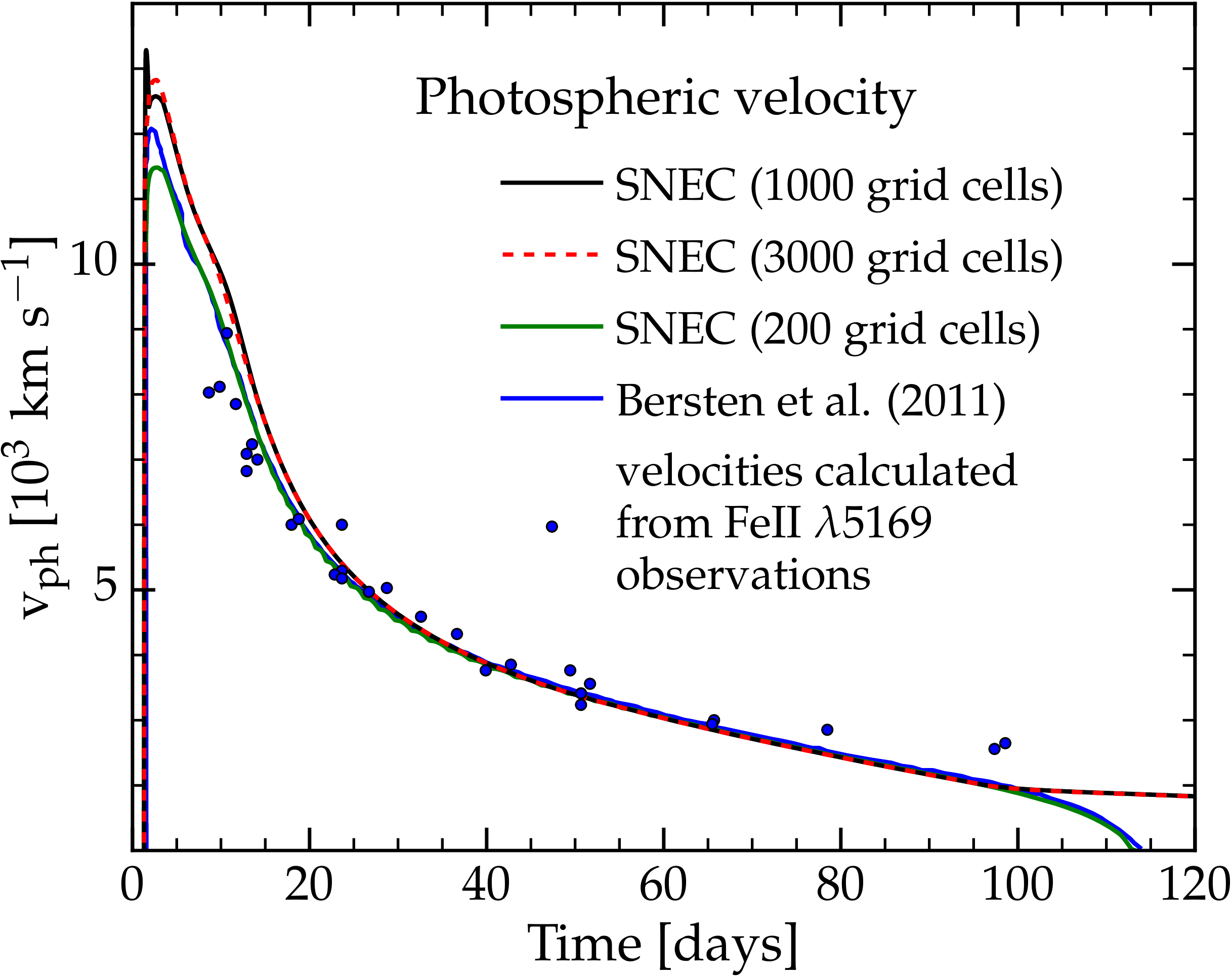}
\end{center}
\caption{Left panel: A comparison of bolometric light curves for SN
  1999em.  The black and red lines show \texttt{SNEC} light curves
  generated with $1000$ and $3000$ grid cells, respectively. The blue
  circles show observational data and the blue curve is the light
  curve of \cite{bersten:11}, both taken from their
  Figure~5. The green graph shows the closest
    light curve we could obtain to the one of \citet{bersten:11} (see
    text for details). The dashed line gives the total power input
  due to radioactive decay. Right panel:
  Photospheric velocity, calculated with
    \texttt{SNEC} for the same model and resolutions as in the left
    panel. The blue circles show observational data and the blue
    graph shows the result of \cite{bersten:11}, both taken from their
    Figure~6.  Note, that we do not take the opacity floor into
    account when determining the location of the photosphere, as
    described in Section \ref{sec:opac}.} \label{1999em_lum}
\end{figure}

The left panel of the Figure~\ref{1999em_lum}
  shows bolometric light curves generated with \texttt{SNEC} and the
  light curve taken from Figure~5 of \citet{bersten:11}, together with
  the observational data for SN 1999em.  The green light curve was
  generated with \texttt{SNEC} using the original $200$-cell grid
  setup of the model of \cite{bersten:11}. Two other \texttt{SNEC}
  light curves, which were generated from the same model mapped onto
  grids with $1000$ and $3000$ cells, disagree with the $200$ grid
  cells curve at early times as well as during the transition between
  plateau and $^{56}\mathrm{Ni}$ tail.  We have identified two
  things that may be key contributors to these differences.

\begin{figure}
\begin{center}
\includegraphics[width=0.46\textwidth]{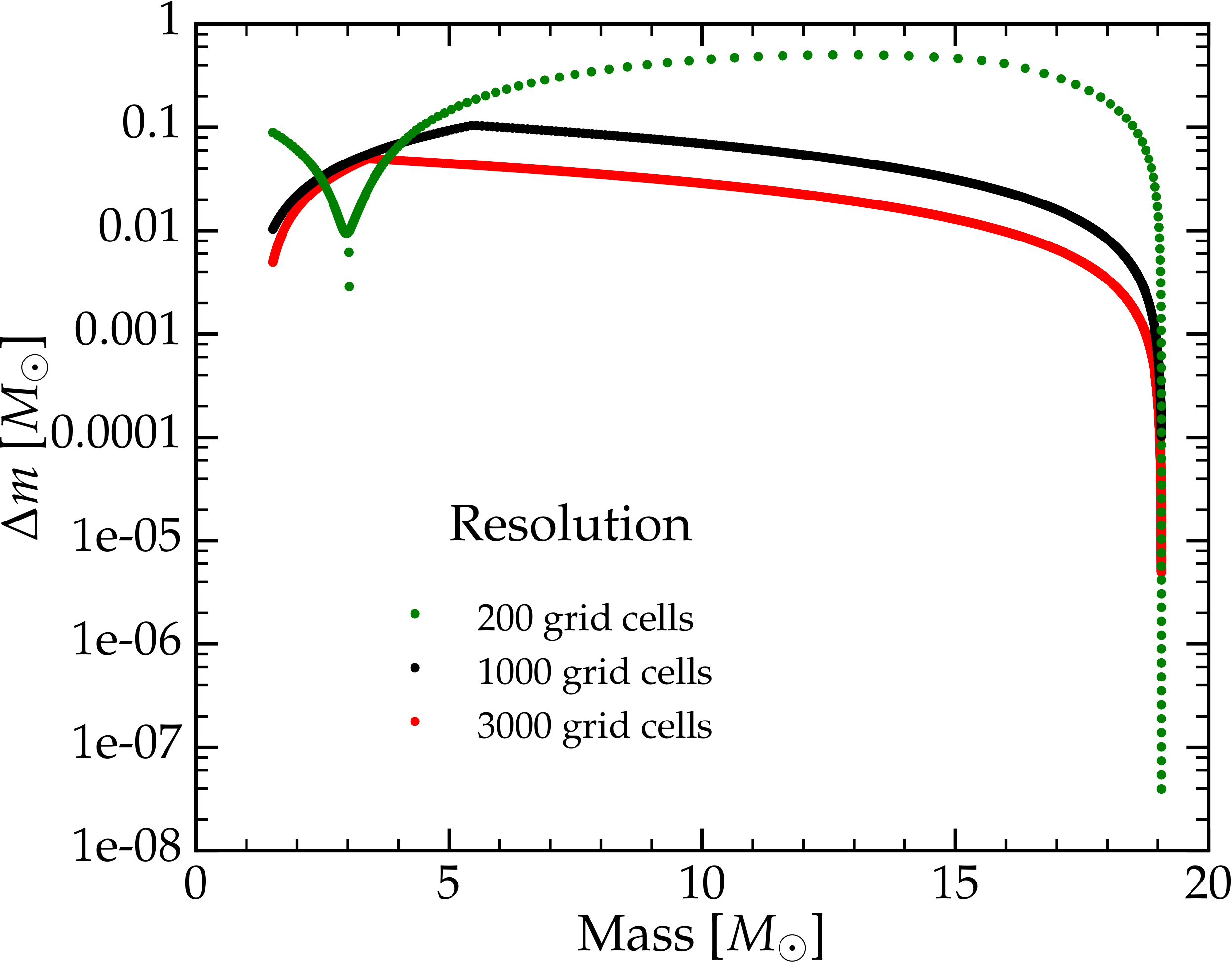}
\end{center}
              \caption{Mass resolution as a function of mass coordinate for the
              $200$, $1000$ and $3000$ grid cells simulations from 
              Figure~\ref{1999em_lum}. The grid having $200$ cells is the
              original grid of the model, which we received from the
              authors of \cite{bersten:11}.}\label{resolutions}
\end{figure}

First, the differences at early times before the
  plateau phase may be explained by differences in resolution and the
  resolution gradient between the $200$-cell grid on the one hand, and
  the $1000$-cell and $3000$-cell grids on the other hand.  To further
  illustrate the differences in resolution, we plot in
  Figure~\ref{resolutions} the mass resolution as a function of
  enclosed mass for the three grids. Note that during the first 50
  days of the light curve, the photosphere moves inwards from a mass
  coordinate of $19\,M_{\odot}$ (the total mass of the model) to a
  mass coordinate of $16\,M_{\odot}$.  The $16-19\,M_\odot$ region is
  precisely where Figure~\ref{resolutions} shows very large
  differences in resolution between the different grids.  The original
  $200$-cell grid of the double polytropic model of \citet{bersten:11}
  has the finest resolution near the surface of the star, but very
  rapidly changes to the coarsest resolution in the bulk of the
  model. In fact, we encountered numerical difficulties exploding the
  $200$-cell model. We find that we have to use an outer boundary
  condition that is different from that described in Section
  \ref{sec:code} of this paper: for numerical stability, we find that
  we have to impose zero temperature at the surface of the star
  instead of constant luminosity in the two outermost grid points. The
  numerical evolutions and light curves from the $1000$-cell and
  $3000$-cell runs, however, have negligible dependence on the switch
  between these two boundary conditions. Note that the light curves
  from the $1000$-cell and $3000$-cell runs lie on top of each other,
  demonstrating that our \texttt{SNEC} results are numerically
  converged.

Second, the difference in the transition from the
  plateau to the $^{56}\mathrm{Ni}$ tail may be explained by the
  difference in opacities we use for the $200$-cell run on one hand,
  and $1000$-cell and $3000$-cell runs on the other hand. For the
  $200$-cell run, we use opacity tables that are different from those
  described in Section \ref{sec:opac}: In the
  low-temperature region ($10^{2.7}\,\mathrm{K} < T <
  10^{4.5}\,\mathrm{K}$) where OPAL tables are not available, we
  employ the Ferguson~\emph{et al.}~tables \citep{ferguson:05} for all
  densities, temperatures, and compositions.  These tables depend on
  hydrogen mass fraction, density, and temperature, and otherwise
  assume simply rescaled solar composition. Hence, we ignore the
  dependence of the opacity on variations of the carbon and oxygen
  mass fractions. We find that this approach is
  essential for reproducing the light curve of
  \cite{bersten:11}. Note, however, that these authors do not
  explicitly state how they treat the opacity in low-temperature,
  Carbon/Oxygen rich regions.  The \texttt{SNEC} light curves
  generated with 1000 and 3000 grid cells use our standard opacities
  as described in Section \ref{sec:opac}.  As mentioned in that
  section, the choice of opacity in the low-temperature, high carbon
  and oxygen mass fractions region weakly influences the evolution of
  the system, because most of these opacities lie below the opacity
  floor. However, it does impact the position of the photosphere
  during the transition between the plateau and the $^{56}$Ni tail, as
  can be seen in Figure~\ref{1999em_lum}. We emphasize that the green
  light curve in the left panel of Figure~\ref{1999em_lum} cannot be
  reproduced with the standard version of \texttt{SNEC}. We provide it
  here only to demonstrate how closely we can approach the results of
  \citet{bersten:11}.  Nevertheless, this shows that overall we find
  reasonable agreement between the light curves shown in
  Figure~\ref{1999em_lum} and have an understanding of the small
  differences.

We also compare the velocity evolution
calculated with the two codes.
The right panel of Figure~\ref{1999em_lum} shows the expansion
velocity at the position of the photosphere, calculated with
\texttt{SNEC} using the same progenitor model and the grid
resolutions as in the left panel of Figure~\ref{1999em_lum}. The observational
data are taken from \citet{bersten:11}. As in the
left panel of Figure~\ref{1999em_lum}, the $200$ grid cells curve
shows the best agreement with \citet{bersten:11}.

As an additional point, it is important to understand whether our treatment
applies to light curves at the earliest times.
The typical grid setup in \texttt{SNEC} focuses
resolution close to the surface of the progenitor star in order to
ensure the photospheric region is well resolved as early as possible
(as discussed in  Section \ref{sec:explosion_setup}). A detailed
analysis of the very early light curve around shock breakout was
carried out by \citet{ensman:92}. In particular, these authors
compared the results obtained with a two-temperature
radiation-hydrodynamics treatment and with a one-temperature
flux-limited equilibrium diffusion treatment similar to
\texttt{SNEC}. They concluded that the two approaches give consistent
results for light curve and photospheric temperature, provided the
surface grid resolution is sufficiently fine so that at the onset of
shock breakout a few tens of grid cells are covering the optically
thin region outside the photosphere. In our \texttt{SNEC} models, we
find that this level of photosphere resolution is not practical for
the large model grid we are considering. However, even with our
standard gridding, we find that after the first few hours of the
explosion that this is sufficient to resolve the
photosphere and produce a reliable light curve in the cooling phase
after shock breakout.

\begin{figure}
\begin{center}
\includegraphics[width=0.46\textwidth]{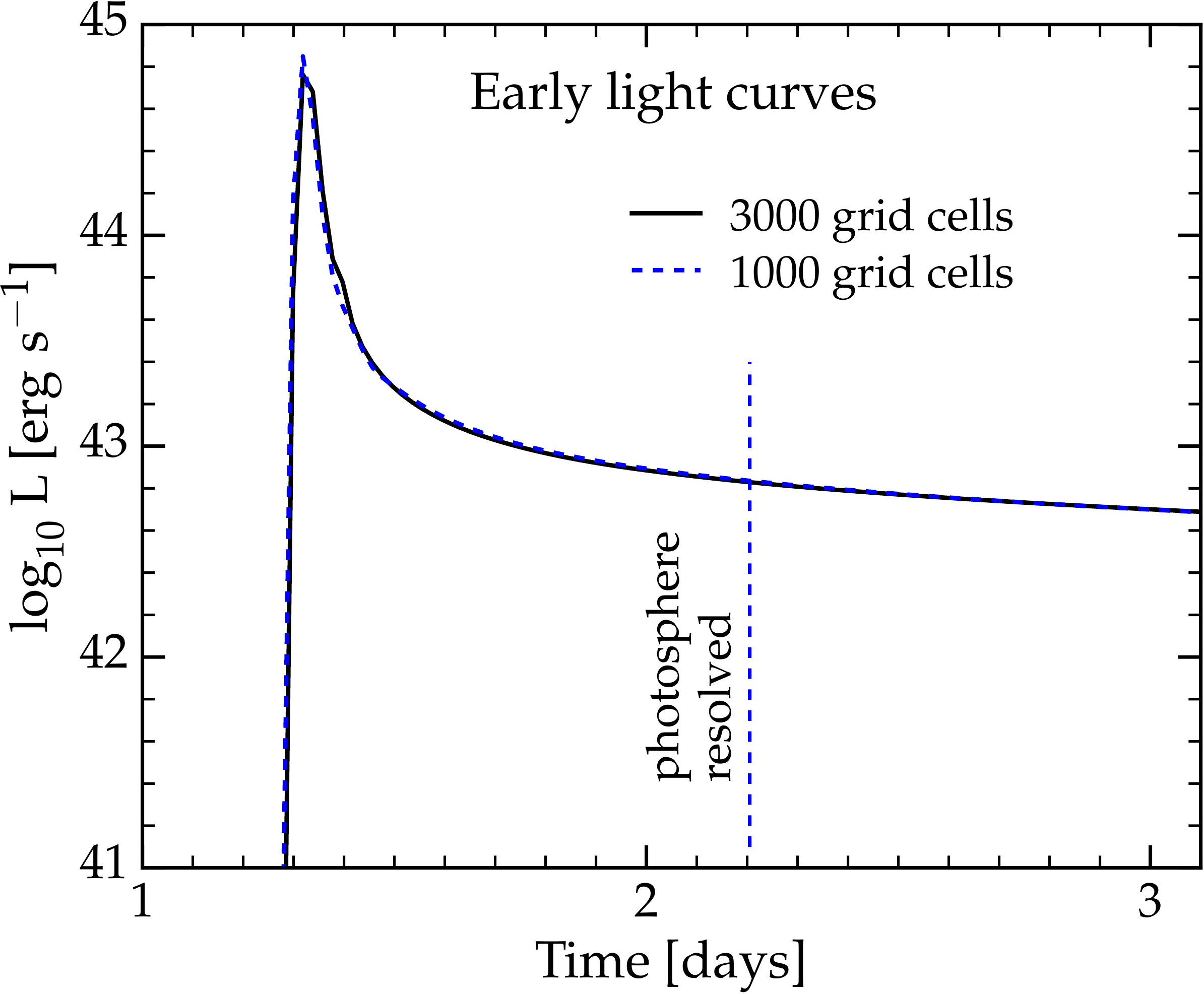}
\end{center}
              \caption{Bolometric light curves around the time of shock breakout using the SN 1999em
              progenitor model.  The black
  solid and blue dashed curves are \texttt{SNEC} light curves
  generated with $1000$ and $3000$ cells, respectively. The
  higher-resolution model always resolves the photosphere, and although the
  lower-resolution model does not resolve it until 
  $\sim$2.2 days, it agrees quite well.} \label{early_curve}
\end{figure}

In order to demonstrate this point, in
Figure~\ref{early_curve} we plot the early light curves around shock breakout
obtained with \texttt{SNEC} using $1000$ and $3000$ grid cells and the
same grid setup as described above for the same progenitor model as in
Figure~\ref{1999em_lum}. The model with $3000$ grid cells always has
at least a few grid cells above the photosphere, whereas in the
lower-resolution model, the photosphere is remains in the outermost grid
cell until it becomes resolved only about one day after breakout. Once this
happens, the results of the $1000$ and $3000$ agree nearly
perfectly. But even at earlier times, the two light curves agree
surprisingly well. This can be understood from the fact that already
shortly after shock breakout the light curve is determined by the
diffusion of light from the shock heated envelope and should not
depend anymore on the resolution of the surface layers.

\section{Comparison with Dessart~{et~al.}\ (2013)}
\label{Ap2}

As an additional comparison, we consider
the light curves obtained by \citet{dessart:13} from a
\texttt{MESA} presupernova progenitor. This is exploded with the
Lagrangian 1D hydrodynamics code \texttt{V1D}
\citep{livne:93,dessart:10}, which is then mapped at day $10$ to their non-LTE
radiative transfer code \texttt{CMFGEN} \citep{hillier:12} that
assumes homologous expansion. \texttt{CMFGEN} provides a more
detailed treatment of radiative transfer than \texttt{SNEC}.
For our specific comparison, we choose model m15Mdot, which was evolved
with enhanced mass loss. The set of parameters provided in
\citet{dessart:13} is sufficient to generate a \texttt{MESA}
progenitor model with similar but not exactly the same as Dessart~et~al.'s
model m15Mdot. The presupernova hydrogen-rich envelope mass in our
version of model m15Mdot is $M_\mathrm{H} \sim 8.13\,M_\odot$ and the
model's radius is $R \sim 789\,R_\odot$. \cite{dessart:13} find
$M_\mathrm{H} \sim 7.72\,\,M_\odot$ and $R \sim 776\,R_\odot$. we
attribute the differences to our version to the different release
versions of \texttt{MESA} used by \cite{dessart:13} and us. Given the
differences in the models (and in the codes), we do not expect
perfect agreement.

Our \texttt{SNEC} explosions are triggered by a thermal bomb, while
\cite{dessart:13} use a piston.  We tried to match the explosion
parameters used by \citet{dessart:13} (given in their Table~2) as
closely as possible.  In particular, we excised the inner
$1.5\,M_{\odot}$ in agreement with the location of the piston in the
models of \citet{dessart:13}. We choose the explosion energy in such a
way that the final total energy of the model is
$1.28\times10^{51}\,\mathrm{erg}$. To achieve this, we inject
{$1.44\times10^{51}\,\mathrm{erg}$} into
the innermost $0.02\,M_{\odot}$ over a time of $0.1\,\mathrm{sec}$.
In addition, we include $0.081\,M_{\odot}$ of $^{56}\mathrm{Ni}$,
initially uniformly distributed in the mass coordinate range $1.5 -
3.5\,M_{\odot}$. We apply boxcar smoothing, as for the other
models from our study. Although we do not know the
precise distribution of $^{56}\mathrm{Ni}$ in the model of
\citet{dessart:13}, the authors indicate that it was not strongly
mixed outwards.  We used the same prescription for the opacity floor
as for all our other models in this study.

\begin{figure}
\begin{center}
\includegraphics[width=0.5\textwidth]{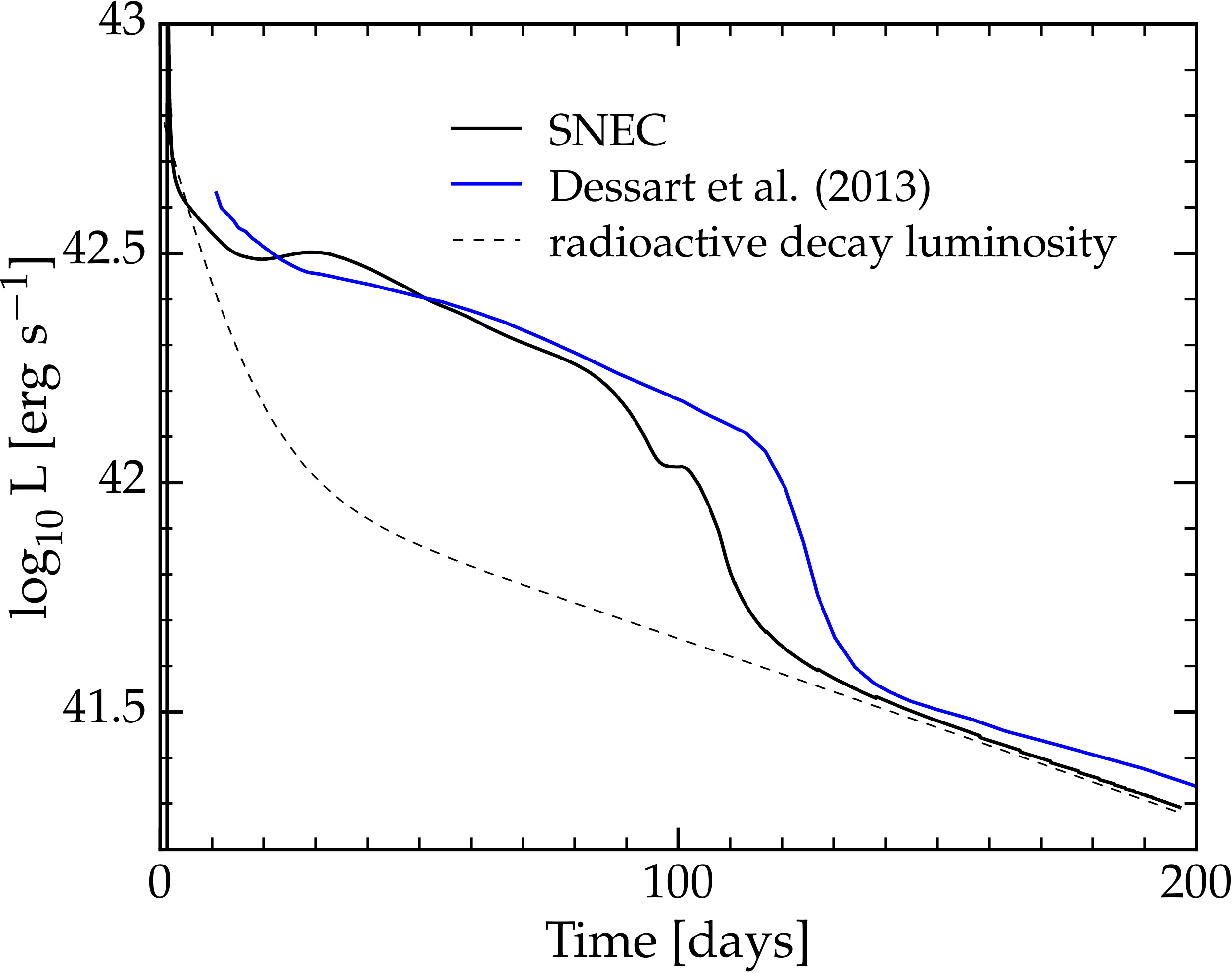}
\end{center}
\caption{Comparison of the bolometric light curve of model m15Mdot of
  \citep[blue line,][]{dessart:13} with the light curve that we obtain with
  \texttt{SNEC} for a similar progenitor model, the same explosion
  energy, and the same initial $^{56}$Ni mass (black line). The
  Dessart~et~al.\ light curve is generated with their \texttt{CMFGEN}
  non-LTE radiative transfer code that assumes homologous expansion
  \cite{hillier:12}.} \label{lum_dessart}
\end{figure}

Figure~\ref{lum_dessart} compares our \texttt{SNEC} light curve with
the light curve obtained by \citet{dessart:13} and shown in their
Figure~7.  We find encouraging agreement between the two light curves
in the first $\sim$$80$ days, as well as after day 140 (in the nebular
phase). The Dessart~et~al.\ plateau
is $10-20\,\mathrm{days}$ longer than predicted by \texttt{SNEC}.
There are a couple of possible reasons for this difference to consider.

On one hand, the two progenitor models are not identical and the
Dessart~et~al.\ model has less mass (by $\sim$$0.4\,M_\odot$) in its
hydrogen-rich envelope than our model. However, the higher
hydrogen-rich mass should result in a somewhat longer plateau
(e.g.~\citealt{young:04,kasen:09}) in our model, which is not what we
find in Figure~\ref{lum_dessart}. Also, one notes from
Figure~\ref{lum_dessart} that the two light curves would agree much
better if the Dessart~et~al.\ light curve was shifted back in time by
$\sim$$5\,\mathrm{days}$. It is not clear what would cause such a
shift. However, we point out that we find in our 
\texttt{SNEC}
calculations that at day $10$ the expansion is not yet homologous and
more internal energy has yet to be converted into kinetic energy of
expansion (this point was previously discussed by
\citealt{dessart:11}). Nevertheless, Dessart~et~al., who assume
homologous expansion, map to their radiative transfer code already at
day ten.  The consequences of such an early mapping should be
explored.

On the other hand (Dessart, \emph{private communication}), the
differences may be caused by \texttt{SNEC}'s radiation-hydrodynamics
treatment, the assumption of LTE, and in the opacities we use (see
Section \ref{SNEC}).  \texttt{CMFGEN} solves the
  time-dependent radiative transfer equations, resolving up to $2000$
  and more so called super-levels in the frequency domain, which
  typically represent $5000-10000$ atomic levels
  \citep[see][]{hillier:12}. In optically thin regions,
  \texttt{CMFGEN} uses line opacities and emissivities instead of the
  Rosseland mean values, used by \texttt{SNEC}.  This provides a better
  description of the effects of iron group line blanketing. In order
to further isolate potential culprits we have carried out experiments
in which we varied the explosion energy, the degree of
$^{56}\mathrm{Ni}$ mixing, and the opacity floor. None of these tests
produce light curves that are substantially closer to the
\cite{dessart:13} light curves than what is shown in
Figure~\ref{lum_dessart}.

\end{document}